\documentclass[journal]{vgtc}                     %

\onlineid{1333}

\vgtccategory{Research}

\vgtcpapertype{application/design study}

\renewcommand{\manuscriptnotetxt}{© 2023 IEEE.
This is the author's version of the article that has been accepted in IEEE VIS 2023.
Personal use of this material is permitted.
Permission from IEEE must be obtained for all other uses, in any current or future media, including reprinting/republishing this material for advertising or promotional purposes, creating new collective works, for resale or redistribution to servers or lists, or reuse of any copyrighted component of this work in other works.
}

\PassOptionsToPackage{svgnames}{xcolor}
\usepackage{xcolor}
\usepackage{xspace, xpunctuate}
\usepackage{listings}  %
\usepackage{enumitem}  %
\usepackage{multirow}
\definecolor{keywordcolor}{rgb}{0.56, 0.13, 0.00}
\definecolor{ndkeywordcolor}{rgb}{0.05, 0.46, 0.17}
\definecolor{commentcolor}{rgb}{0.41, 0.64, 0.70}
\definecolor{stringcolor}{rgb}{0.25, 0.44, 0.63}
\lstdefinelanguage{TypeScript}{
  keywords={typeof, new, true, false, catch, function, return, null, catch, switch, var, if, in, while, do, else, case, break, boolean},
  morekeywords={[2]{class, export, throw, implements, import, this}},
  identifierstyle=\color{black},
  sensitive=false,
  comment=[l]{//},
  morecomment=[s]{/*}{*/},
  commentstyle=\color{commentcolor}\ttfamily,
  stringstyle=\color{stringcolor}\ttfamily,
  morestring=[b]',
  morestring=[b]"
}
\newcommand{\inlinecode}[1]{\texttt{\detokenize{#1}}}
\newcommand{\inlinecodesmall}[1]{{\fontsize{8}{12}\texttt{\detokenize{#1}}}}

\setlist{noitemsep,parsep=0pt,partopsep=0pt}

\newcommand{\ea}{{et~al\xperiod}\xspace}

\usepackage[group-separator={$,$}]{siunitx}

\newcommand{\numOldVis}{\num{13511}\xspace}
\newcommand{\numOldVisShort}{13K\xspace}
\newcommand{\numVis}{\num{19571}\xspace}
\newcommand{\numDataSources}{seven\xspace}

\newcommand{\nDavidRumseyMapCollectionImages}{25563}
\newcommand{\numDavidRumseyMapCollectionImages}{\num{\nDavidRumseyMapCollectionImages}\xspace}
\newcommand{\nTotalImages}{370883}
\newcommand{\numTotalImages}{\num{\nTotalImages}\xspace}

\newcommand{\githubLink}{https://github.com/oldvis}

\newcommand{\dbname}{\textit}

\title{OldVisOnline: Curating a Dataset of Historical Visualizations}
\author{
    \authororcid{Yu~Zhang}{0000-0002-9035-0463},
    \authororcid{Ruike~Jiang}{0000-0002-5001-5310},
    \authororcid{Liwenhan~Xie}{0000-0002-2601-6313},
    \authororcid{Yuheng~Zhao}{0000-0003-1573-8772},\\
    \authororcid{Can~Liu}{0000-0002-1175-0734},
    Tianhong~Ding,
    \authororcid{Siming~Chen}{0000-0002-2690-3588},
    and \authororcid{Xiaoru~Yuan}{0000-0002-7233-980X}
}

\authorfooter{
    \item
    Yu~Zhang is with the Department of Computer Science, University of Oxford.
    E-mail: yuzhang94@outlook.com.

    \item
    Ruike~Jiang, Can~Liu, and Xiaoru~Yuan (corresponding author) are with the Key Laboratory of Machine Perception (Ministry of Education) and the School of Intelligence Science and Technology, Peking University.
    E-mail: \{jiangrk | can.liu | xiaoru.yuan\}@pku.edu.cn.

    \item
    Yuheng~Zhao and Siming~Chen (corresponding author) are with the School of Data Science, Fudan University.
    E-mail: \{yuhengzhao | simingchen\}@fudan.edu.cn.

    \item
    Liwenhan~Xie is with the Department of Computer Science and Engineering, The Hong Kong University of Science and Technology.
    E-mail: liwenhan.xie@connect.ust.hk.

    \item
    Tianhong~Ding is with the Fundamental Software Innovation Lab, Huawei Technologies.
    The research is conducted as an independent researcher and not an outcome of work at Huawei.
    E-mail: dingtianhong@huawei.com.
}

\abstract{
   With the increasing adoption of digitization, more and more historical visualizations created hundreds of years ago are accessible in digital libraries online.
It provides a unique opportunity for visualization and history research.
Meanwhile, there is no large-scale digital collection dedicated to historical visualizations.
The visualizations are scattered in various collections, which hinders retrieval.
In this study, we curate the first large-scale dataset dedicated to historical visualizations.
Our dataset comprises \numOldVisShort historical visualization images with corresponding processed metadata from \numDataSources digital libraries.
In curating the dataset, we propose a workflow to scrape and process heterogeneous metadata.
We develop a semi-automatic labeling approach to distinguish visualizations from other artifacts.
Our dataset can be accessed with OldVisOnline, a system we have built to browse and label historical visualizations.
We discuss our vision of usage scenarios and research opportunities with our dataset, such as textual criticism for historical visualizations.
Drawing upon our experience, we summarize recommendations for future efforts to improve our dataset.

}


\graphicspath{{fig/}{figure/}{picture/}{image/}{./}} %

\usepackage{tabu}                      %
\usepackage{booktabs}                  %
\usepackage{lipsum}                    %
\usepackage{mwe}                       %

\usepackage{mathptmx}                  %

\begin{document}

\ifx\hidemain\undefined
   \maketitle

   \section{Introduction}
\label{sec:introduction}

History has left us with a heritage of visualizations, such as John Snow's map that points to the root cause of the cholera outbreak~\cite{Snow1855Mode} and Charles Joseph Minard's map of Napoleon's Russian campaign that illustrates the losses in the retreat~\cite{Minard1869Carte}.
By examining these legacies, we can gain invaluable insights into how people have communicated and interpreted data.
The wit in the design of historical visualizations has served as guidance for contemporary visualizations~\cite{Tufte1983Visual}.
As with other artifacts, visualizations are records of the thoughts and culture of the time, where we can peek at the social context, artistic norms, and the level of technological advancement in the past.

There have been cumulative efforts gathering historical visualizations and exploiting the wisdom underneath.
For instance, \textit{The Book of Trees}~\cite{Lima2014Book} examines historical practices of encoding hierarchical data.
\textit{History of Information Graphics}~\cite{Rendgen2019History} describes historical infographics spanning various fields such as astronomy and zoology.
The milestones project provides an online gallery of significant visualization inventions in history~\cite{Friendly2001Milestones}.
They serve as excellent resources to inspire and communicate findings on the history of visualization.
However, existing historical visualization collections fall short in scale and diversity.
Myriads of historical visualizations remain scattered across digital libraries and have not been paid sufficient attention in the research community.
Our current knowledge of historical visualization is built on limited instances~\cite{Kosara2016Empire}, whereas an overarching view is missing.

Continuing these prior efforts in consolidating knowledge in data visualization, we are motivated to gather the scattered historical visualizations and assemble a digital library featuring visualizations.
By curating such a collection, we aspire to amplify the impact of historical visualizations.
We aim to center them as an integral part of our understanding, not only of visualization but also history.
In doing so, we address two challenges.
Firstly, digital libraries apply various metadata schemas and use different vocabularies in describing their collections.
It is necessary to develop a consistent and unified approach to scraping and organizing visualizations.
Secondly, visualizations are sparsely distributed in digital libraries, which warrants an effective filter to remove irrelevant artifacts.
The large volume of visualizations necessitates a level of automation in labeling.

This paper presents the first large-scale dataset of diverse historical visualizations.
Our dataset comprises \numOldVisShort visualization images published before 1950 and corresponding metadata with 17 dimensions.
We collect data from \numDataSources digital libraries, such as \dbname{David Rumsey Map Collection}~\cite{RumseyDavid} and \dbname{Gallica}~\cite{NLFGallica}.
Reflecting on our experience curating the dataset, we formulate a workflow to collect, clean, and label visualizations from various data sources.
We highlight that historical visualizations, as an underexplored resource combining text, image, and data, can shed light on studies in history and visualization.
We envision many usage scenarios of the dataset for history and visualization researchers, such as textual criticism and design space analysis.

In summary, this paper has the following contributions:

\begin{itemize}[leftmargin=3.5mm]
      \item We contribute the first dedicated large-scale historical visualization dataset with \numOldVisShort images and corresponding metadata.
            Users can explore it through the image browser we develop.

      \item We propose a dataset construction workflow to collect, clean, and label visualization images.

      \item We discuss in-depth the usage scenarios and research opportunities facilitated by our dataset.
\end{itemize}

The source code, documentation, and dataset are available at \url{\githubLink}.

   \section{Background \& Related Work}

This section briefly introduces the background of using historical visualization in science history research and reviews the literature on datasets of visualizations.

\subsection{Visualization in Science History}

Visualizations are widely adopted in various scientific domains.
Some historical visualizations are regarded as milestones in corresponding scientific domains, such as John Snow's cholera map for epidemiology.
They are thus important records for science history studies and have attracted much attention.
For example, Robinson overviews the life and works of Charles Joseph Minard and presents a chronological listing of Minard's thematic maps~\cite{Robinson1967Thematic}.
Friendly and Denis trace the invention and development of scatterplots in the history of science~\cite{Friendly2005Early}.
Friendly et al. investigate the background of Michael Florent van Langren's 1644 chart showing estimations of the longitude between Toledo and Rome, sometimes regarded as the first known statistical chart~\cite{Friendly2010First}.

Despite the interest in studying historical visualizations from a historical perspective, the lack of dedicated collections of historical visualizations poses an obstacle.
Visualizations are usually sparsely scattered in historical documents and are easily overlooked.
For example, John Snow's \textit{On the Mode of Communication of Cholera}~\cite{Snow1855Mode}, where the famous cholera map was published, contains only two visualizations among around two hundred pages.

Furthermore, existing digital libraries typically lack filters to enable users to retrieve visualizations besides maps.
The lack of support in digital libraries impedes a holistic inspection of historical visualizations.
In response to the call to expand the history of data visualization~\cite{Klein2022What} and avoid over-relying on a small number of classic visualizations~\cite{Kosara2016Empire}, we develop a large-scale historical visualization dataset.
Our work also aligns with the suggestion to promote visualizations into a first-class medium of information~\cite{Wu2022AI4VIS}.

Our dataset focuses on gathering historical visualizations, i.e., visualizations created long ago.
A related effort is Old Maps Online~\cite{KTGOld}, which gathers historical geographic maps.
To our best knowledge, Friendly and Denis' milestones project is the only prior effort on gathering the scattered treasure of various types of historical visualizations~\cite{Friendly2001Milestones}.
They focus on milestones and the firsts in the history of visualization and have manually collected around 400 images.
We build a large-scale historical visualization dataset to provide a rich resource for science history and visualization researchers.
Note that we focus on gathering ``old'' visualizations rather than visualizations of historical datasets like the ones reviewed by Windhager et al.~\cite{Windhager2019Visualization}.

\subsection{Dataset of Visualizations}

Practitioners and researchers have contributed many datasets of contemporary visualizations, typically image datasets.
Such visualization datasets have been exhaustively reviewed by Liu et al.~\cite{Liu2022Visualization} and Ye et al.~\cite{Ye2022VISAtlas}.
The following review focuses on usage scenarios discussed in prior work on visualization datasets.

Large datasets of visualizations enable the discovery of visualization design and usage patterns~\cite{Lee2018Viziometrics, Chen2021Composition, Deng2022Revisiting, Lin2023DMiner}.
For instance, using a collection of 41K visualizations from major venues where developers publish visualizations, Battle et al. discover differences in the chart type preferences among developers using different toolkits~\cite{Battle2018Beagle}.
Using 13K infographics obtained from design websites, Lu \ea identified 12 design patterns of linkages among graphical elements to convey information~\cite{Lu2020Exploring}.
The underlying patterns of visualization datasets also contribute to model training for automatic chart analysis~\cite{Wu2022AI4VIS, Chen2023State}.

Additionally, visualization datasets are real-world resources for empirical studies.
For instance, Borkin \ea conducted crowdsourcing studies to understand what visual features in visualizations contribute to better memorability~\cite{Borkin2013What}, recognition, and recall~\cite{Borkin2016Memorability} based on a dataset with over 5K visualizations scraped from the web.
Hu et al. proposed the VizNet~\cite{Hu2019VizNet} repository comprising datasets of web-based visualizations.
They demonstrate using the datasets to replicate a prior study~\cite{Kim2018Assessing} for evaluating the effectiveness of visual encodings.

Moreover, visualization datasets facilitate overviewing, communicating, and educating visualization techniques.
Some survey papers provide online browsers to exemplify the design space of visualization techniques such as tree~\cite{Schulz2011Treevis.net}, geospatial network~\cite{Schoettler2021Visualizing}, text~\cite{Kucher2015Text}, and data-GIFs~\cite{Shu2020What}.
There has been increasing interest in curating visualization images that appeared in publications at visualization venues such as IEEE VIS and IEEE TVCG.
Examples include ViStory~\cite{Zeng2021VIStory}, VIS30K~\cite{Chen2021VIS30K}, and VisImages~\cite{Deng2023VisImages}.
Chen~\ea~\cite{Chen2022Not} proposed a topography to index visualization images in IEEE VIS papers.

In this work, we have constructed a dataset comprising real-world visualizations created before the computer era.
Our dataset supports usage scenarios specific to or particularly interesting for historical visualizations.
As detailed in \cref{sec:usage-scenarios}, such scenarios include textual criticism, historical data extraction, design space analysis, inspiration for contemporary design, metadata analysis, and metadata research.

With the availability of large-scale visualization datasets, there have been investigations to facilitate efficient explorations of such datasets.
VISAtlas~\cite{Ye2022VISAtlas} projects visualizations onto a 2D plane based on their embeddings to provide an overview and support comparisons between datasets.
Thumbnail design~\cite{Oppermann2021VizSnippets} and search engine designs~\cite{Chen2015DiagramFlyer, Hoque2020Searching} have been explored for information retrieval in visualization collections.

To support the exploration of our historical visualization dataset, we develop the OldVisOnline system.
Like existing literature search and analysis systems, such as SurVis~\cite{Beck2016Visual} and KeyVis~\cite{Isenberg2017Visualization}, OldVisOnline supports metadata searching and filtering.
The image gallery of OldVisOnline follows the design of SurVis~\cite{Beck2016Visual}.
Compared with these prior systems, OldVisOnline additionally comprises data labeling functions.

   \begin{figure*}[htb]
    \centering
    \includegraphics[width=\linewidth]{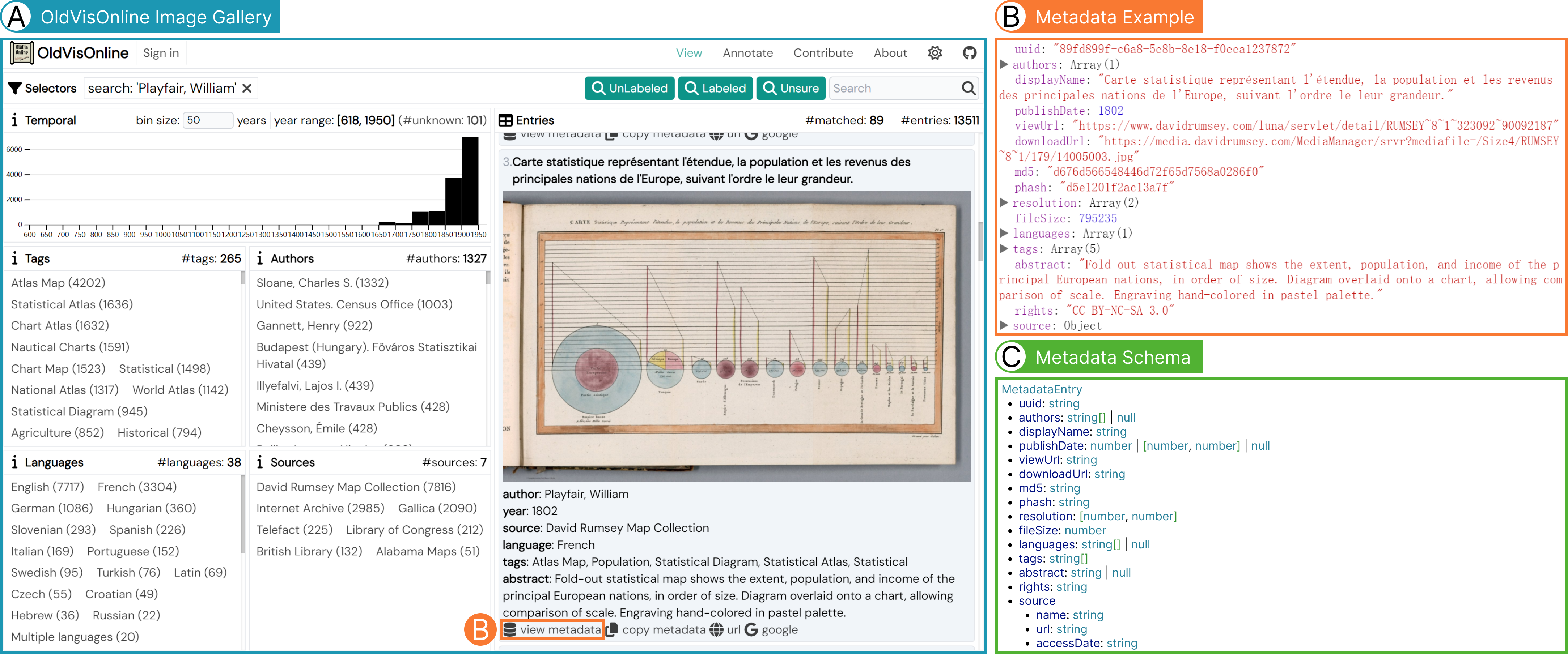}
    \caption{
        \textbf{Browsing historical visualizations in OldVisOnline's image gallery interface:}
        (A) The image gallery interface of OldVisOnline contains images and corresponding metadata of \numOldVisShort historical visualizations published before 1950 that we curate from various data sources.
        The user can search and filter metadata fields of historical visualizations.
        (B) The metadata of a French edition of William Playfair's chart published in 1802 as an example~\cite{Playfair1801Statistical}.
        (C) Our design of historical visualizations' metadata schema.
    }
    \label{fig:online-gallery}
\end{figure*}

\section{Dataset Construction Workflow}
\label{sec:dataset}

This section introduces our workflow for constructing the historical visualization dataset.
We clarify the scope and sources of our collection, elaborate on the process of collecting and filtering data, and discuss considerations and rationales.

\subsection{Data Sources}
\label{sec:data-sources}

We have used seven digital libraries and collections as our data sources, as listed in \cref{table:data-sources}.
We include \dbname{David Rumsey Map Collection}~\cite{RumseyDavid}, \dbname{Internet Archive}~\cite{Kahle1996Internet}, \dbname{Gallica}~\cite{NLFGallica}, \dbname{Library of Congress}~\cite{LibraryCongressLibrary}, and \dbname{British Library}
\footnote{
\dbname{British Library} provides multiple galleries including \href{https://www.bl.uk/collection-items}{\dbname{Collection Items}} and \href{https://imagesonline.bl.uk}{\dbname{Images Online}}.
We use \dbname{British Library} to refer to the union of the two galleries.
}~\cite{BritishLibrary1973British}
because they are well-known massive digital libraries, some of which are authoritative national libraries, thus likely holding more collections of historical visualizations.
\dbname{Telefact}~\cite{Modley1938Telefact} and \dbname{Alabama Maps}~\cite{CRLAlabama} are included based on our previous experience in identifying historical visualizations in them.
We use multiple digital libraries instead of relying on a general-purpose image search engine, such as Google Images, with three considerations:

\begin{itemize}[leftmargin=3.5mm]
    \item \textbf{Data availability:}
          Most digital libraries do not expose their collections to search engines.
          Many images are only available through these digital libraries.

    \item \textbf{Metadata richness:}
          Images in digital libraries are accompanied by metadata, which are crucial for further analysis.
          Such metadata are absent for entries returned by search engines.

    \item \textbf{Data quality:}
          The data quality of digital libraries is higher than that of search engines.
          The images indexed by digital libraries are usually high-resolution and deduplicated.
\end{itemize}

\subsubsection{Inclusion Criteria}

To make clear what should be included in our historical visualization dataset, we set the following inclusion criteria to address the two constraints, ``historical'' and ``visualization'':

\begin{itemize}[leftmargin=3.5mm]
    \item \textbf{Publish date:}
          By ``historical'', we require the visualization to be published before 1950.
          Using computerized graphics design toolkits gradually became the most common approach for creating visualizations since Sutherland proposed Sketchpad~\cite{Sutherland1963Sketchpad}.
          Setting the publish date threshold to 1950 allows us to focus on the visualizations created before the computer age.

    \item \textbf{Content:}
          By ``visualization'', we refer to the common definition of it as ``visual representations of datasets''~\cite{Munzner2014Visualization}.
          While maps are generally regarded as visualizations of spatial data, we exclude maps from our collection except when the maps contain information layers that encode non-spatial data.
          Historical maps are extensively studied in historical geography.
          It is easy to obtain hundreds of thousands of historical maps from dedicated collections such as Old Maps Online~\cite{KTGOld}.
          Thus, we ignore conventional maps to focus on other types of historical visualizations, which are much less explored.
\end{itemize}

We acknowledge that users may have different understandings of the temporal constraint set by ``historical''.
Thus, when constructing the dataset, we provide another edition of the dataset without the publish date constraint, as shown in \cref{table:data-sources}.
Using this dataset edition, the users can create their editions of the historical visualization dataset by applying customized publish date filters if needed.

\subsubsection{Search Strategies}
\label{sec:search-strategies}

Most data sources have massive collections and are not dedicated to historical visualizations.
Thus, gathering all the images from the data sources is impractical and unnecessary.
To retrieve historical visualizations from data sources that support searching, we mainly search with two types of keywords:

\begin{itemize}[leftmargin=3.5mm]
    \item \textbf{Synonyms of ``visualization'':} examples include ``visualization'', ``chart'', and ``diagram''.

    \item \textbf{Names of visualization authors:} examples include ``John Snow'', ``Charles Joseph Minard'', and ``{\'{E}}mile Cheysson''.

\end{itemize}

We try the search strategies for each data source that supports searching and keep the search keywords that return historical visualizations.
For data sources with advanced search functions such as publication date filtering, we tailor our search queries toward their search grammar.
\iflabelexists{sec:unify-raw-metadata}{\Cref{sec:unify-raw-metadata}}{The supplemental material}
shows the queries that we adopted after trial and error.
We deduplicate the query results for the data sources where we conduct multiple queries.
The number of images we obtain from each data source is shown in \cref{table:data-sources}.

\subsubsection{Metadata Schema}
\label{sec:metadata-schema}

Four of the data sources (\dbname{David Rumsey Map Collection}, \dbname{Internet Archive}, \dbname{Gallica}, and \dbname{Library of Congress}) provide APIs for querying images and metadata through HTTP GET requests.
However, none of them share the same API or return the same data structure for the image metadata.
Thus, we develop a separate data collection script for each data source.
We normalize the metadata from different sources into the same schema as shown in \cref{fig:online-gallery}(C).
The details of each metadata field are listed below:

\begin{itemize}[leftmargin=3.5mm]
    \item \textbf{uuid:}
          The universally unique identifier (UUID) of the metadata entry.
          For reproducibility, the UUID should be generated with UUID version 5 and use the internal identifier of the corresponding data source as the seed.

    \item \textbf{authors:}
          The authors of the visualization.

    \item \textbf{displayName:}
          A short title of the visualization for display.

    \item \textbf{publishDate:}
          The time point or range of publish date.

    \item \textbf{downloadUrl:}
          The URL where the image can be downloaded with an HTTP GET request.

    \item \textbf{viewUrl:}
          The URL where the image can be viewed in a browser.

    \item \textbf{md5:}
          The MD5 hash of the image.

    \item \textbf{phash:}
          The perceptual hash of the image.

    \item \textbf{resolution:}
          The [width, height] of the image in pixels.

    \item \textbf{fileSize:}
          The storage size of the image in bytes.

    \item \textbf{languages:}
          The languages used in the visualization (in ISO 639-3 codes~\cite{IOS2007Codes}).

    \item \textbf{tags:}
          The index terms of the visualization for searching.

    \item \textbf{abstract:}
          A brief description of the visualization.

    \item \textbf{rights:}
          The copyright status of the image.

    \item \textbf{source.name:}
          The name of the data source.

    \item \textbf{source.url:}
          The URL where the raw metadata were obtained.

    \item \textbf{source.accessDate:}
          The time when the raw metadata were obtained (in ISO 8601 format~\cite{IOS2019Date}).
\end{itemize}

As there is no consensus on the data structure returned from APIs of digital libraries, we consult the BibTeX file format in designing this data structure.
We include bibliography information such as \inlinecode{authors}, \inlinecode{displayName} (i.e., title), and \inlinecode{publishDate}.
As such, we support exporting the metadata in BibTeX format to facilitate interoperability with bibliographic management software.
We also add attributes tailored for images.
\inlinecode{md5} and \inlinecode{phash} are stored to support image content checksum and comparison.
\inlinecode{resolution} and \inlinecode{fileSize} are stored to support size-related filtering of the dataset.
\iflabelexists{sec:unify-raw-metadata}{\Cref{sec:unify-raw-metadata}}{The supplemental material}
describes the technical details of unifying the raw metadata from each data source into this data structure.

\subsection{Data Labeling and Filtering}
\label{sec:data-labeling-and-filtering}

Not all the images in the query results obtained from the data sources are historical visualizations.
To curate a collection of historical visualizations, we filter the non-visualization images with the following semi-automatic data labeling and label quality assurance processes.

\begin{figure}[htb]
    \centering
    \includegraphics[width=\linewidth]{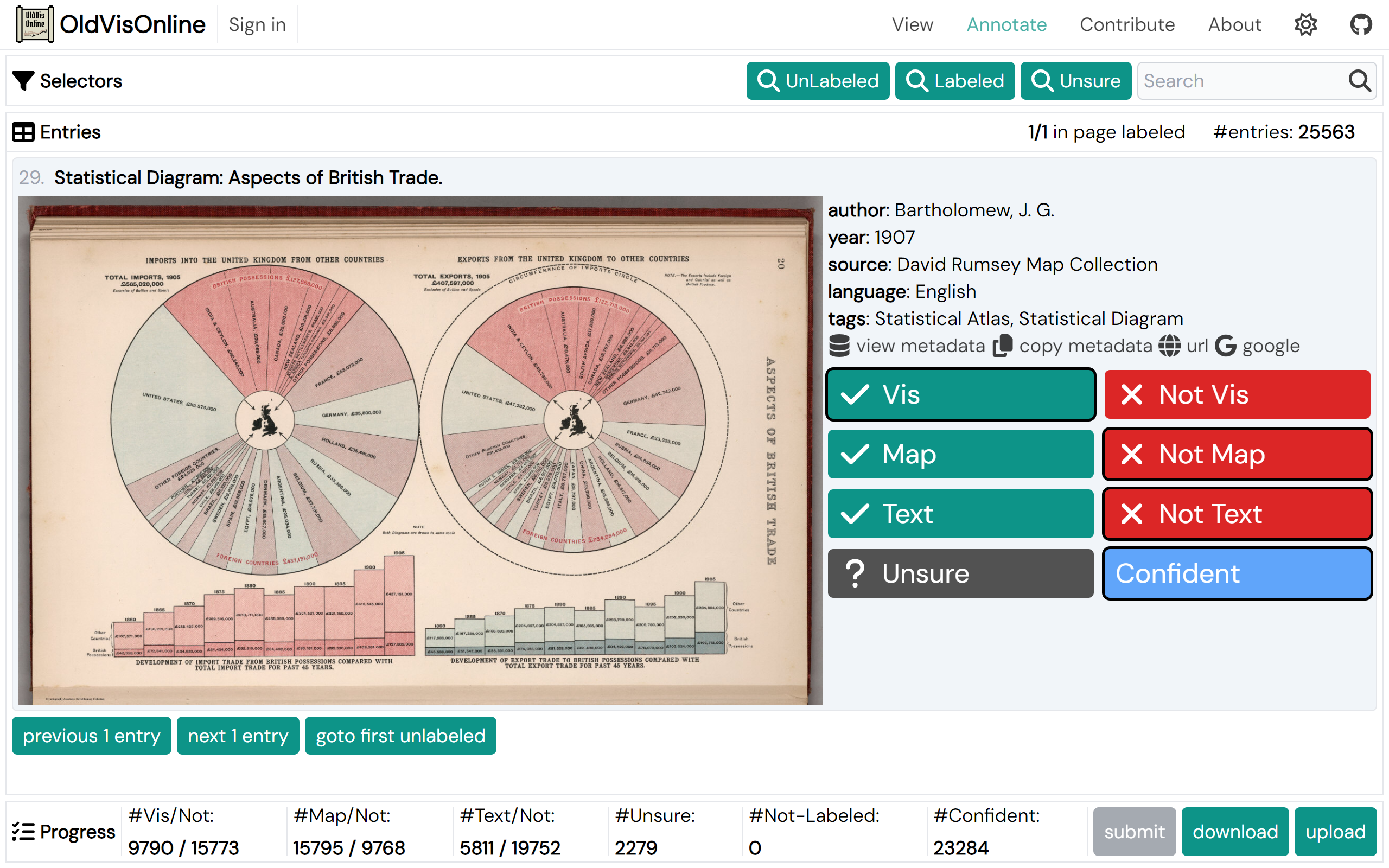}
    \caption{
        \textbf{Categorizing images with OldVisOnline's image classification interface:} The interface shows one image each time, together with its metadata.
        The annotator can classify each image into \textit{Vis}/\textit{NotVis}, \textit{Map}/\textit{NotMap}, and \textit{Text}/\textit{NotText} through button clicking.
        The annotator can also assign \textit{Unsure}/\textit{Confident} to denote the annotation confidence.
    }
    \label{fig:image-classification-interface}
\end{figure}

\subsubsection{Semi-Automatic Labeling}
\label{sec:semi-automatic-labeling}

In the semi-automatic labeling process, we first manually label the images collected from \dbname{David Rumsey Map Collection}, \dbname{British Library Images Online}, \dbname{Telefact}, and \dbname{Alabama Maps}.
Then, we train classifiers to predict the labels of images in the remaining data sources.
The following describes each stage of this semi-automatic labeling process.

\textbf{Manual labeling:}
Using the labeling interface shown in \cref{fig:image-classification-interface}, we manually assign three binary classification labels to each image:
\begin{itemize}[leftmargin=3.5mm]
    \item \textit{Vis}/\textit{NotVis:} contains a visualization or not.
    \item \textit{Map}/\textit{NotMap:} contains a map or not.
    \item \textit{Text}/\textit{NotText:} has over half of the area occupied by text or not.
\end{itemize}

The manual labeling is carried out by one of the authors.
Labeling the \numDavidRumseyMapCollectionImages images in \dbname{David Rumsey Map Collection} took around 160 hours.
Through labeling, we identify that \num{9790} images contain visualizations, \num{15795} images contain maps, and \num{5811} images have over half of the area occupied by text.
Note that the labels are not mutually exclusive.
An image can be \textit{Vis}, \textit{Map}, and \textit{Text} at the same time.
Likewise, manual labeling is conducted on the images in \dbname{British Library Images Online}, \dbname{Telefact}, and \dbname{Alabama Maps}.

\textbf{Model training:}
Using the annotations for \dbname{David Rumsey Map Collection}, we train three binary classifiers:
\begin{itemize}[leftmargin=3.5mm]
    \item $C_{text - vis - map}$: classify whether an image is \textit{Text}, \textit{NotVis}, and \textit{NotMap} at the same time.
    \item $C_{vis + map}$: classify whether an image is \textit{Vis} or \textit{Map}.
    \item $C_{vis}$: classify whether an image is \textit{Vis}.
\end{itemize}

$C_{text - vis - map}$ detects text pages, a common type of non-visualization images in the data sources.
The text pages are usually easy to differentiate from visualization images.
Applying $C_{text - vis - map}$ to filter images allows us to effectively eliminate text pages that are common and easy cases of non-visualization images.

\begin{table}[!htbp]
    \caption{
        The label distribution and model performance for the three binary classification tasks: $text - vis - map$, $vis + map$, and $vis$.
        ``Positive'' and ``Negative'' refers to the number of data points labeled positive and negative in \dbname{David Rumsey Map Collection}.
        ``Accuracy'' and ``F1'' refers to the accuracy and F1 score of the models trained with 80\% of the labeled dataset and validated on the remaining 20\%.
    }
    \label{table:binary-classification}
    \centering
    \scriptsize
    \begin{tabular}{lrrrr}
        \toprule
        \textbf{Task}      & \textbf{Positive} & \textbf{Negative} & \textbf{Accuracy} & \textbf{F1} \\
        \midrule
        $text - vis - map$ & \num{5147}        & \num{20416}       & 97.4\%            & 0.936       \\
        $vis + map$        & \num{18020}       & \num{7543}        & 97.3\%            & 0.981       \\
        $vis$              & \num{9790}        & \num{15773}       & 92.2\%            & 0.897       \\
        \bottomrule
    \end{tabular}
\end{table}

$C_{vis + map}$ detects maps and visualizations.
In ideal situations with accurate model predictions, $C_{vis}$ should be directly applied to detect visualizations.
In practice, maps are hard to differentiate from visualization, as can be seen from differences between the performances of $C_{vis + map}$ and $C_{vis}$ in \cref{table:binary-classification}.
Thus, we use a separate classifier $C_{vis + map}$ to detect both visualization and maps and leave the challenging task of differentiating visualizations and maps to the last step.

The three classifiers are trained with the same hyperparameter setup.
We fine-tune VGG-16~\cite{Simonyan2014Very} pre-trained on ImageNet-1K~\cite{Deng2009ImageNet}.
We train the model with two stages as practiced in prior work~\cite{Krizhevsky2012ImageNet,Deng2023VisImages}.
In the first stage, we freeze the weights of convolutional layers and train the classification head for 6 epochs.
In the second stage, we train all the weights for 6 epochs.
We optimize the binary cross entropy loss using stochastic gradient descent with the initial learning rate being $1e^{-3}$.

We utilize the manual annotations on \dbname{David Rumsey Map Collection} to train the three classifiers.
\Cref{table:binary-classification} shows the number of positive and negative data points as well as model performances for the three tasks.
When evaluating the model performances, we train the model with 80\% of the labeled dataset and validate with the remaining 20\%.
When using the model in practice, to make the most from the annotations, we train the three classifiers with 100\% of the labeled dataset.

\begin{figure}[htb]
    \centering
    \includegraphics[width=\linewidth]{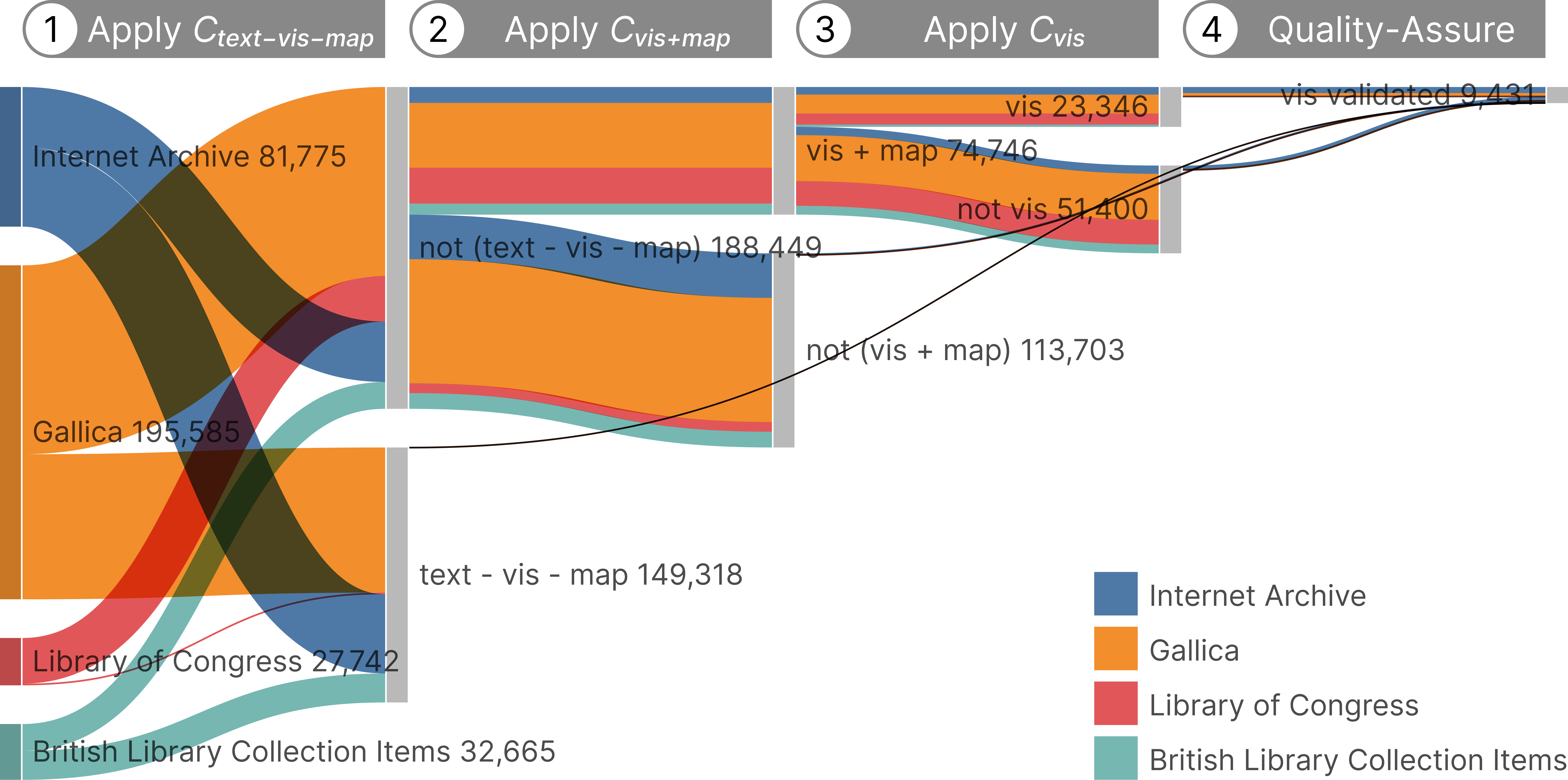}
    \caption{
        The number of images discarded and kept in each step of the filtering and quality assurance processes.
    }
    \label{fig:data-filtering}
\end{figure}

\textbf{Prediction and filtering:}
With trained classifiers $C_{text - vis - map}$, $C_{vis + map}$, and $C_{vis}$, we carry out a series of filtering steps on the \num{337767} unlabeled images in \dbname{Internet Archive}, \dbname{Gallica}, \dbname{Library of Congress}, and \dbname{British Library Collection Items}.
\Cref{fig:data-filtering} shows the number of images discarded and kept in each step.

\begin{itemize}[leftmargin=3.5mm]
    \item \textbf{Step 1: Apply $C_{text - vis - map}$ to all the images and discard the ones classified as positive.}
    Among all the images, 44.21\% (\num{149318}/\num{337767}) are positive (i.e., are text pages) and discarded (marked as $text - vis - map$ in \cref{fig:data-filtering}).
    
    \item \textbf{Step 2: Apply $C_{vis + map}$ to the remaining images and discard the ones classified as negative.}
    Among the remaining images that $C_{text - vis - map}$ classified as negative, 60.34\% (\num{113703}/\num{188448}) are negative and discarded (marked as $not\ (vis + map)$ in \cref{fig:data-filtering}).
    
    \item \textbf{Step 3: Apply $C_{vis}$ to the remaining images and discard the ones classified as negative.}
    Among the remaining images that $C_{vis + map}$ classified as positive, 68.77\% (\num{51400}/\num{74746}) are negative and discarded (marked as $not\ vis$ in \cref{fig:data-filtering}).
\end{itemize}

Through the filtering steps, among all the images, the models predict 6.91\% (\num{23346}/\num{337767}) as potential visualizations (marked as $vis$ in \cref{fig:data-filtering}).
The predictions are not directly used as definitive results.
In the following, we introduce the quality assurance of the predictions.

\begin{table}[!htbp]
    \caption{
        The number of identified visualization images and quality assurance time costs for the four disjoint data subsets derived in the filtering steps.
        The union of the subsets forms the set of \num{337767} unlabeled images in  \dbname{Internet Archive}, \dbname{Gallica}, \dbname{Library of Congress}, and \dbname{British Library Collection Items}.
        ``Vis\%'' refers to the percentage of visualizations images among all the images in the data subset.
        ``Time'' refers to the quality assurance time costs measured in minutes.
        ``Speed'' refers to the number of images quality-assured per second.
    }
    \label{table:quality-assurance-summary}
    \centering
    \scriptsize
    \scalebox{0.97}{
        \begin{tabular}{lrrrrr}
            \toprule
            \textbf{Data Subset}            & \textbf{\#Images} & \textbf{\#Vis} & \textbf{Vis\%} & \textbf{Time} & \textbf{Speed (\#/s)} \\
            \midrule
            $\{text - vis - map\}$          & \num{149318}      & 118            & 0.08\%         & 128           & 19.5                  \\
            $\{\mathrm{not}\ (vis + map)\}$ & \num{113703}      & 1014           & 0.89\%         & 206           & 9.2                   \\
            $\{\mathrm{not}\ vis\}$         & \num{51400}       & 2639           & 5.13\%         & 219           & 3.9                   \\
            $\{vis\}$                       & \num{23346}       & 5660           & 24.24\%        & 289           & 1.3                   \\
            \midrule
            \textbf{Total}                  & \num{337767}      & 9431           & 2.79\%         & 842           & 6.7                   \\
            \bottomrule
        \end{tabular}
    }
\end{table}

\begin{table*}[!htbp]
    \centering
    \scriptsize
    \caption{
        The performance of multi-label classifiers with different hyper-parameter setups.
        ``LR'' refers to the learning rate.
        ``Loss'' refers to the binary cross entropy loss.
        We highlight the best performance under each measure in \textbf{bold}.
    }
    \setlength{\aboverulesep}{0.5pt}
    \setlength{\belowrulesep}{0.5pt}
    \label{table:multilabel-classification}
    \begin{tabular}{c|c|c|cccc|cccc}
        \toprule
        \multirow{2}[0]{*}{\textbf{Model}}                & \multirow{2}[0]{*}{\textbf{LR}} & \multirow{2}[0]{*}{\textbf{Loss}} & \multicolumn{4}{c|}{\textbf{Accuracy}} & \multicolumn{4}{c}{\textbf{F1}}                                                                                                                                                                \\\cline{4-11}
                                                          &                                 &                                   & \multicolumn{1}{c}{All}                & \multicolumn{1}{c}{Vis}         & \multicolumn{1}{c}{Map} & \multicolumn{1}{c|}{Text} & \multicolumn{1}{c}{All} & \multicolumn{1}{c}{Vis} & \multicolumn{1}{c}{Map} & \multicolumn{1}{c}{Text} \\
        \midrule
        \multirow{3}{*}{ConvNeXt-T~\cite{Liu2022ConvNet}} & 1e-2                            & 0.622                             & 67.0\%                                 & 61.9\%                          & 61.7\%                  & 77.2\%                    & 0.254                   & 0                       & 0.763                   & 0                        \\
                                                          & 1e-3                            & 0.125                             & 95.1\%                                 & 91.6\%                          & \textbf{97.1\%}         & 96.7\%                    & 0.932                   & 0.891                   & 0.976                   & 0.928                    \\
                                                          & 1e-4                            & 0.167                             & 93.3\%                                 & 88.8\%                          & 95.5\%                  & 95.7\%                    & 0.906                   & 0.848                   & 0.963                   & 0.908                    \\\hline
        \multirow{3}{*}{ResNet-50~\cite{He2016Deep}}      & 1e-2                            & 0.173                             & \textbf{95.4\%}                        & \textbf{92.2\%}                 & \textbf{97.1\%}         & 96.8\%                    & 0.934                   & \textbf{0.897}          & \textbf{0.977}          & 0.929                    \\
                                                          & 1e-3                            & 0.181                             & 94.0\%                                 & 90.1\%                          & 95.8\%                  & 96.2\%                    & 0.916                   & 0.864                   & 0.966                   & 0.918                    \\
                                                          & 1e-4                            & 0.271                             & 89.1\%                                 & 82.2\%                          & 93.0\%                  & 92.2\%                    & 0.841                   & 0.744                   & 0.943                   & 0.835                    \\\hline
        \multirow{3}{*}{VGG-16~\cite{Simonyan2014Very}}   & 1e-2                            & 0.150                             & 95.0\%                                 & 91.4\%                          & 96.8\%                  & 96.8\%                    & 0.929                   & 0.882                   & 0.974                   & 0.931                    \\
                                                          & 1e-3                            & 0.135                             & 95.3\%                                 & 91.8\%                          & 96.9\%                  & \textbf{97.2\%}           & \textbf{0.936}          & 0.891                   & 0.975                   & \textbf{0.941}           \\
                                                          & 1e-4                            & 0.164                             & 93.6\%                                 & 88.8\%                          & 95.6\%                  & 96.3\%                    & 0.913                   & 0.857                   & 0.965                   & 0.917                    \\
        \bottomrule
    \end{tabular}
\end{table*}

\subsubsection{Quality Assurance of Model Predictions}

The author responsible for manual annotation in the manual labeling stage screened all the model predictions and corrected misclassifications.
The screening was conducted in the default file explorer of Windows 11.
The images were sorted by file size before the screening.
The three filtering steps create four disjoint data subsets:
\begin{itemize}[leftmargin=3.5mm]
    \item $\{text - vis - map\}$: the images classified positive by $C_{text - vis - map}$ in the first step.
    \item $\{\mathrm{not}\ (vis + map)\}$: the images classified negative by $C_{vis + map}$ in the second step.
    \item $\{\mathrm{not}\ vis\}$: the images classified negative by $C_{vis}$ in the third step.
    \item $\{vis\}$: the images classified positive by $C_{vis}$ in the third step.
\end{itemize}

\Cref{table:quality-assurance-summary} shows the percentage of visualization images and quality assurance time cost for the four data subsets.
In total, \num{9431} visualization images are identified and validated from the data subsets in the screening (marked as $vis\ validated$ in \cref{fig:data-filtering}).

The percentage of visualizations increases through the filtering steps.
Among the images classified as text pages by $C_{text - vis - map}$ in the first step, only 0.08\% are visualizations.
By comparison, among the images classified as visualizations by $C_{vis}$ in the third step, 24.24\% are visualizations.
The filtering steps managed to distill visualizations from the images.

Compared with the manual labeling stage, the quality assurance process is efficient.
The model predictions reduce human effort as the annotator only needs to make corrections infrequently instead of annotating from scratch~\cite{Zhang2023Simulation}.
The quality assurance of model predictions on \num{337767} images took 842 minutes.
We observe that the speed of quality assurance decreases with the filtering step.
As shown in \cref{table:quality-assurance-summary}, for the images discarded in the first step (i.e., images in $\{text - vis - map\}$), the speed of manual quality assurance was 19.5 images per second.
By comparison, for the images kept in the final step (i.e., images in $\{vis\}$), the speed of manual quality assurance was 1.3 images per second.
The speed decrease may be due to increased image content complexity and image label ambiguity through the filtering steps.

We quality-assure model predictions on all the images to obtain high-quality annotations and to evaluate the classifiers' effectiveness in filtering visualization images.
Depending on the usage scenario, future maintainers of our dataset or similar datasets may not quality-assure all the model predictions.
For example, when the goal is not to obtain as many visualization images as possible but to obtain visualization images as efficiently as possible, the dataset curator may quality-assure only the images classified as visualizations (i.e., images in $\{vis\}$) and discard all the other images directly without checking.

\subsubsection{Considerations}

The following discusses the considerations and rationales for details in data labeling and filtering.

\textbf{On rule-based filtering:}
Compared with training classifiers, an alternative approach is to program rule-based filters.
We have trialed this approach but gave up because the metadata sometimes lack enough information to support the classification.
For example, a visualization image may be a page of a book.
It can be hard to differentiate the visualization image from the remaining text pages by metadata, as the metadata of all the book pages can be similar.

\textbf{On crowdsourcing versus expert annotation:}
One of the authors was responsible for data labeling and quality assurance of model predictions.
We decided not to crowdsource annotations as the labeling requires expertise.
For example, it can be difficult to differentiate Joseph Priestley's timeline~\cite{Priestley1803Lectures} filled with text from an ordinary text page.
The borderline between ordinary geographic maps and visualizations can also be hard to judge.

\textbf{On improving label consistency:}
The annotator's understanding of what should be counted as visualization may evolve through the labeling process.
This phenomenon, referred to as concept evolution~\cite{Kulesza2014Structured}, may introduce systematic errors to the labels.
To reduce such errors, we set up weekly meetings for ten months to discuss edge cases met in data labeling and as we tried out our online gallery.
The \textit{Unsure} label in the interface shown in \cref{fig:image-classification-interface} marks the edge cases for later inspection.

\textbf{On model hyper-parameter selection:}
Before model training, the hyper-parameters underwent the following selection process.
We tested three model architectures: ResNet-50~\cite{He2016Deep}, VGG-16~\cite{Simonyan2014Very}, and ConvNeXt-T~\cite{Liu2022ConvNet}.
For each architecture, we tested three learning rates: $1e^{-2}$, $1e^{-3}$, and $1e^{-4}$.
For each combination of the model architecture and learning rate, we fine-tuned the model with weights pre-trained on ImageNet-1K~\cite{Deng2009ImageNet} to output three binary classification labels: \textit{Vis}/\textit{NotVis}, \textit{Map}/\textit{NotMap}, and \textit{Text}/\textit{NotText}.
Each model was trained with 80\% of the labeled data points in \dbname{David Rumsey Map Collection}.
\Cref{table:multilabel-classification} shows the model performances.
It can be seen that the top three setups with similar performances are VGG-16 with $LR = 1e^{-3}$, ResNet-50 with $LR = 1e^{-2}$, and ConvNeXt-T with $LR = 1e^{-3}$
Thus, we used VGG-16 with $LR = 1e^{-3}$ in the model training stage.

\textbf{On iterative model training:}
In the quality assurance process, it is possible to iteratively fine-tune the model with the newly quality-assured data labels.
Iterative training may improve model performance and thus reduce the annotator's quality assurance time cost.
We did not iteratively train the model because of the training time cost.
Training the three binary classifiers in \cref{sec:semi-automatic-labeling} with the \numDavidRumseyMapCollectionImages labeled data points in \dbname{David Rumsey Map Collection} took around 6 hours with an NVIDIA GeForce RTX 3070 Laptop GPU.
Assuming that model parameters are fixed, the training time cost is at least linear with the dataset size.
Iteratively fine-tuning the model with the total \numTotalImages data points would take at least $6 \times \frac{\nTotalImages}{\nDavidRumseyMapCollectionImages} \approx 87$ hours.
By comparison, the manual quality assurance for these data points took around 14 hours.
The annotator would need to wait for the model update if the model were iteratively fine-tuned.
Thus, we refrained from retraining the model iteratively.
In the cases where the developer has access to more computation resources or smaller models are used, iterative model training may bring benefits.

\textbf{On active learning:}
Active learning techniques iteratively select unlabeled data points for annotators to label.
They are commonly used in data labeling systems~\cite{Settles2011Closing,Heimerl2012Visual,Kucher2017Active} for two potential advantages:

\begin{itemize}[leftmargin=3.5mm]
    \item \textbf{Fewer required labels:}
          Active learning aims to train a model to achieve higher accuracy with fewer labels~\cite{Settles2009Active}.
          In some cases, the goal of data labeling is not a fully labeled dataset but a trained model.
          Fewer labels and labeling efforts are required in these cases where the goal of data labeling aligns with active learning.

    \item \textbf{Improved default label accuracy:}
          The model and the predictions can be updated with the iteratively obtained labels.
          Active learning may boost the accuracy of machine-predicted labels and thus reduce the annotator's effort to correct misclassifications.
\end{itemize}

We decided not to use active learning as the two advantages are absent in our scenario.
The number of required labels cannot be reduced as we want the dataset to be fully labeled and checked.
As discussed above, we decided not to iteratively train the model because of the time cost.
Thus, the default labels cannot be updated, and their accuracy cannot benefit from active learning.

   \begin{table}[!htbp]
    \caption{
        We obtain \numVis visualizations from seven data sources with \numOldVis published before 1950.
        ``\#Images'' refers to the number of images we obtain from the data sources.
        ``\#Vis'' refers to the number of visualizations we identify after labeling and filtering in~\cref{sec:data-labeling-and-filtering}.
        ``Before 1950'' refers to the number of visualizations published before 1950.
    }
    \label{table:data-sources}
    \centering
    \scriptsize
    \scalebox{0.95}{
        \begin{tabular}{lrrr}
            \toprule
            \textbf{Data Source}                                                                                                                          & \textbf{\#Images}                  & \textbf{\#Vis} & \textbf{Before 1950} \\
            \midrule
            \includegraphics[height=2mm, width=2mm]{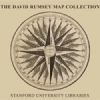} David Rumsey Map Collection~\cite{RumseyDavid} & \numDavidRumseyMapCollectionImages & \num{9790}     & \num{7816}           \\
            \includegraphics[height=2mm, width=2mm]{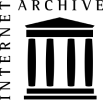} Internet Archive~\cite{Kahle1996Internet}                 & \num{82225}                        & \num{6885}     & \num{2985}           \\
            \includegraphics[height=2mm, width=2mm]{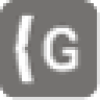} Gallica~\cite{NLFGallica}                                          & \num{195585}                       & \num{2183}     & \num{2090}           \\
            \includegraphics[height=2mm, width=2mm]{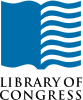} Library of Congress~\cite{LibraryCongressLibrary}      & \num{27742}                        & 299            & 212                  \\
            \includegraphics[height=2mm, width=2mm]{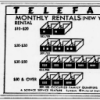} Telefact~\cite{Modley1938Telefact}                                & 264                                & 225            & 225                  \\
            \includegraphics[height=2mm, width=2mm]{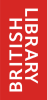} British Library~\cite{BritishLibrary1973British}           & \num{39312}                        & 137            & 132                  \\
            \includegraphics[height=2mm, width=2mm]{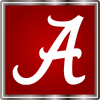} Alabama Maps~\cite{CRLAlabama}                                & 142                                & 52             & 51                   \\
            \midrule
            \textbf{Total}                                                                                                                                & \numTotalImages                    & \numVis        & \numOldVis           \\
            \bottomrule
        \end{tabular}
    }
\end{table}

\section{OldVisOnline}

As shown in \cref{table:data-sources}, through the dataset construction workflow, we obtain \numOldVis visualization images published before 1950, which we regard as historical visualizations.
The visualization images and corresponding metadata are obtained from \numDataSources data sources.
\Cref{fig:online-gallery}(B) shows a metadata entry example.

With the images and metadata, we develop OldVisOnline, a system for browsing and labeling historical visualizations.
OldVisOnline consists of an image gallery interface (\cref{fig:online-gallery}(A)), an image classification interface (\cref{fig:image-classification-interface}), and a fine-grained labeling interface (\cref{fig:fine-grained-labeling-interface}).
The data labeling components of OldVisOnline are adapted from the OneLabeler framework~\cite{Zhang2022OneLabeler}.

OldVisOnline's image gallery supports searching with the metadata.
The image gallery is designed following SurVis~\cite{Beck2016Visual}.
The user can conduct cross-filtering on the \inlinecode{publishDate}, \inlinecode{tags}, \inlinecode{authors}, \inlinecode{languages}, and \inlinecode{source} fields in the corresponding panel on the left of the gallery interface in \cref{fig:online-gallery}(A).
The \emph{Selectors} panel shows the cross-filtering conditions.
The user can also create freeform search queries with the search box on the right of the \emph{Selectors} panels.
The \emph{Entries} panel in \cref{fig:online-gallery}(A) shows the matched entries.
The matched entries are paginated with 20 entries per page.

The panels on the left of the gallery interface (\cref{fig:online-gallery}(A)) show the metadata statistics.
We have extracted 265 index terms for the visualizations.
The visualizations are created by 1327 authors and institutes.
The visualizations are published in 38 languages, as predicted by an n-gram-based text classifier.

   \section{Usage Scenarios}
\label{sec:usage-scenarios}

With the large-scale collection of historical visualizations, we envision multitudinous opportunities for future research.
As with datasets of contemporary visualizations, our historical visualization dataset can serve as a corpus for machine learning~\cite{Deng2023VisImages} and an educational resource~\cite{Chen2021VIS30K}.
Additionally, various usage scenarios are specific to or particularly interesting for historical visualizations.
In this section, we elaborate on such usage scenarios of our dataset.

\subsection{Textual Criticism}
\label{sec:textual-criticism}

Textual criticism is a branch of bibliography and philology that studies variations and errors in different copies of the same historical document.
Our historical visualization dataset and image gallery can serve as a resource and a search engine to help historians conduct textual criticism studies on historical visualizations.
We introduce a case of comparing editions of John Snow's cholera map.

\begin{figure}[htb]
    \centering
    \includegraphics[width=\linewidth]{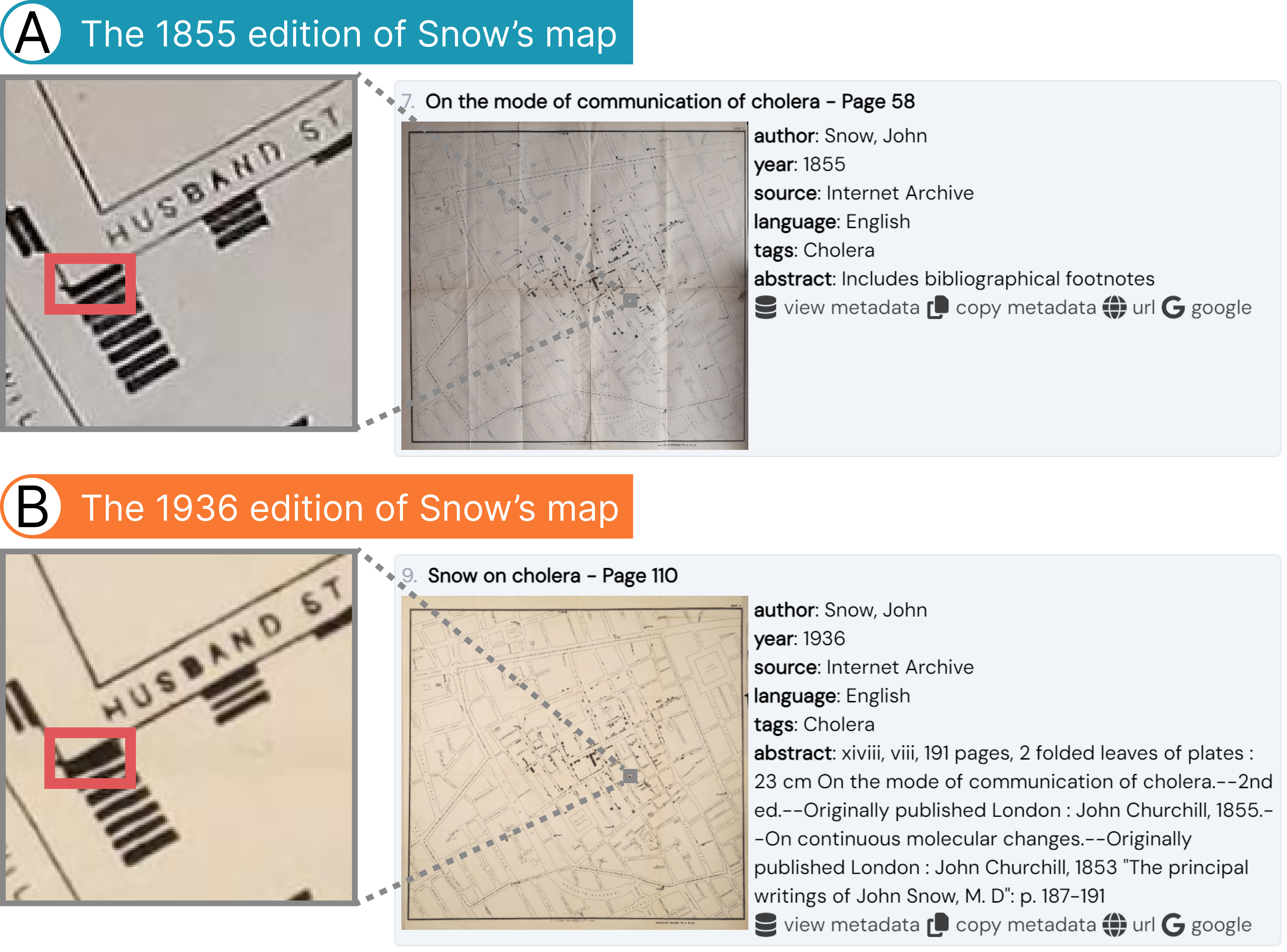}
    \caption{
        \textbf{Comparing two editions of John Snow's cholera map:}
        (A) The \href{https://archive.org/details/b28985266/page/n57}{1855 edition} shows eight rectangular blocks denoting cholera cases at the intersection of \textit{NEW STREET} and \textit{HUSBAND ST}~\cite{Snow1855Mode}.
        (B) The \href{https://archive.org/details/b3134818x_0001/page/n110}{1936 edition} shows seven rectangular blocks at the same location.
        The highlighted block appears to be two blocks conglutinated due to ink diffusion~\cite{Snow1936Snow}.
    }
    \label{fig:textual-criticism}
\end{figure}

After searching ``Snow, John'' in OldVisOnline's image gallery, we identify two editions of the cholera map.
\Cref{fig:textual-criticism}(A) shows the one published in 1855~\cite{Snow1855Mode}.
\Cref{fig:textual-criticism}(B) shows the other published in 1936~\cite{Snow1936Snow}.
Through an initial screening of the two editions, we observe that the thickness of printed letters appears less uniform in the 1936 edition.
Through a literature search, we find that this difference has caused different interpretations of the dataset encoded in the cholera map.
In Zhang et al.'s work on data reconstruction from historical visualizations~\cite{Zhang2021MI3}, they reconstruct data from the cholera map and suggest that the map shows $579$ cholera cases.
By comparison, some other papers~\cite{Koch2004Map, Shiode2015mortality} refer to Dodson and Tobler's dataset~\cite{Dodson1994Snows}, and suggest the map shows $578$ cholera cases.

After plotting Zhang et al.'s dataset and Dodson and Tobler's dataset, we observe that the two datasets differ in the number of cholera cases at the intersection of \textit{NEW STREET} and \textit{HUSBAND ST} in the map, as highlighted in \cref{fig:textual-criticism}.
In the 1855 edition that derived Zhang et al.'s result, eight blocks corresponding to cholera cases are located at the intersection.
In the 1936 edition that derived Dodson and Tobler et al.'s result, seven blocks corresponding to cholera cases are located at the same place.
We suspect the difference may be due to a printing defect in the 1936 edition.
In the 1936 edition, one of the seven blocks is significantly larger than the others, implying that two blocks may be conglutinated and appear as one large block because of ink diffusion.

The first step of textual criticism is to obtain different editions of the same document.
Through this example of identifying discrepancies in two editions of John Snow's cholera map, we demonstrate that different editions of the same historical visualization can be found in our dataset, which fuels the study on textual criticism for historical visualizations.

\subsection{Historical Data Extraction}

Historical visualizations capture valuable historical datasets.
Meanwhile, the original historical data records are often lost over time.
Data extraction is to decode the datasets from the scanned bitmap images of historical visualizations, as shown in \cref{fig:data-extraction}.
The extracted data can be used for further analysis.
Additionally, data extraction enables comparing data encoded in different copies of a historical visualization and thus supports textual criticism.
Prior efforts have been devoted to developing image processing techniques~\cite{Kerle2009Reviving} and interactive machine learning approaches~\cite{Zhang2021MI3} to data extraction.

\begin{figure}[!htbp]
    \centering
    \includegraphics[width=\linewidth]{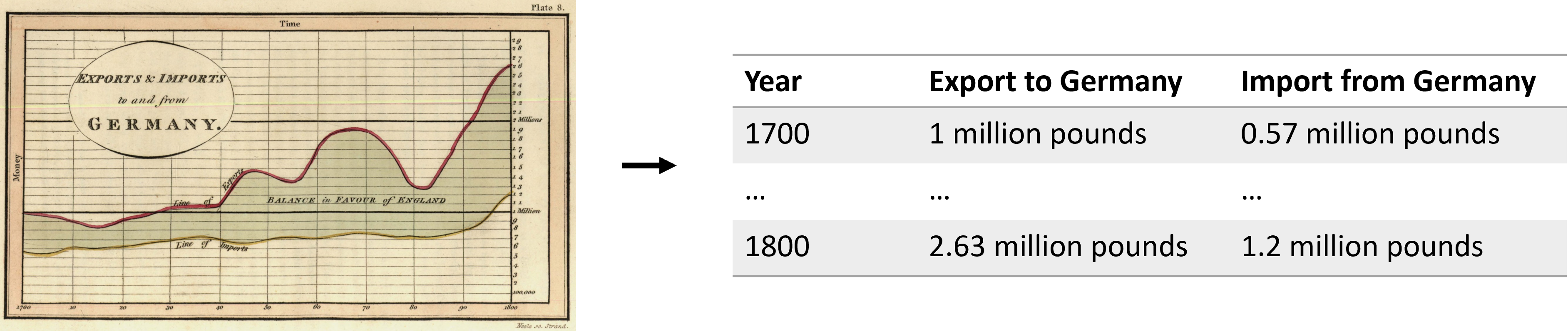}
    \caption{
        \textbf{Data extraction from historical visualizations:}
        Data extraction takes the bitmap image of a visualization as input and outputs a dataset that can reproduce the visualization.
        For example, given a scanned image of Playfair's \href{https://archive.org/details/PLAYFAIRWilliam1801TheCommercialandPoliticalAtlas/page/n69}{line chart} on the left~\cite{Playfair1801Commercial}, the data table on the right shows the outcome of data extraction.
    }
    \label{fig:data-extraction}
\end{figure}

Even when the original dataset is available, comparing them with the extracted data and analyzing the data distortion may still reveal interesting findings.
The distortion may provide evidence of the limitations of the engraving and printing techniques of the times.
It may also reflect the historical visualization authors' values rooted in the political, economic, and cultural background of the times.
Such analysis of data distortion may borrow insights from the analysis of contemporary deceptive visualizations~\cite{Pandey2015How}.

Our dataset provides candidates for historical data extraction.
Moreover, as some of the data extraction approaches utilize supervised learning~\cite{Zhang2021MI3}, our dataset can be enhanced with corresponding labels to serve as training data and benchmark for data extraction models.

\subsection{Design Space Analysis}
\label{sec:design-space-analysis}

Our large-scale historical visualization dataset can be used for analyzing the design space spanned by historical visualizations.
Such analysis may lead to a better understanding of what designs have been explored in history and how visualization techniques and usage patterns have evolved.
Prior work has demonstrated using large-scale contemporary visualization datasets for design space analysis.
Examples include investigating commonly used chart types on the web~\cite{Battle2018Beagle} and analyzing the design patterns of multi-view visualizations~\cite{Chen2021Composition,Deng2022Revisiting}.
Such design space analysis relies on data labels, such as chart type labels~\cite{Battle2018Beagle} and bounding box labels~\cite{Chen2021Composition,Deng2022Revisiting}.

\begin{figure}[!htbp]
    \centering
    \includegraphics[width=\linewidth]{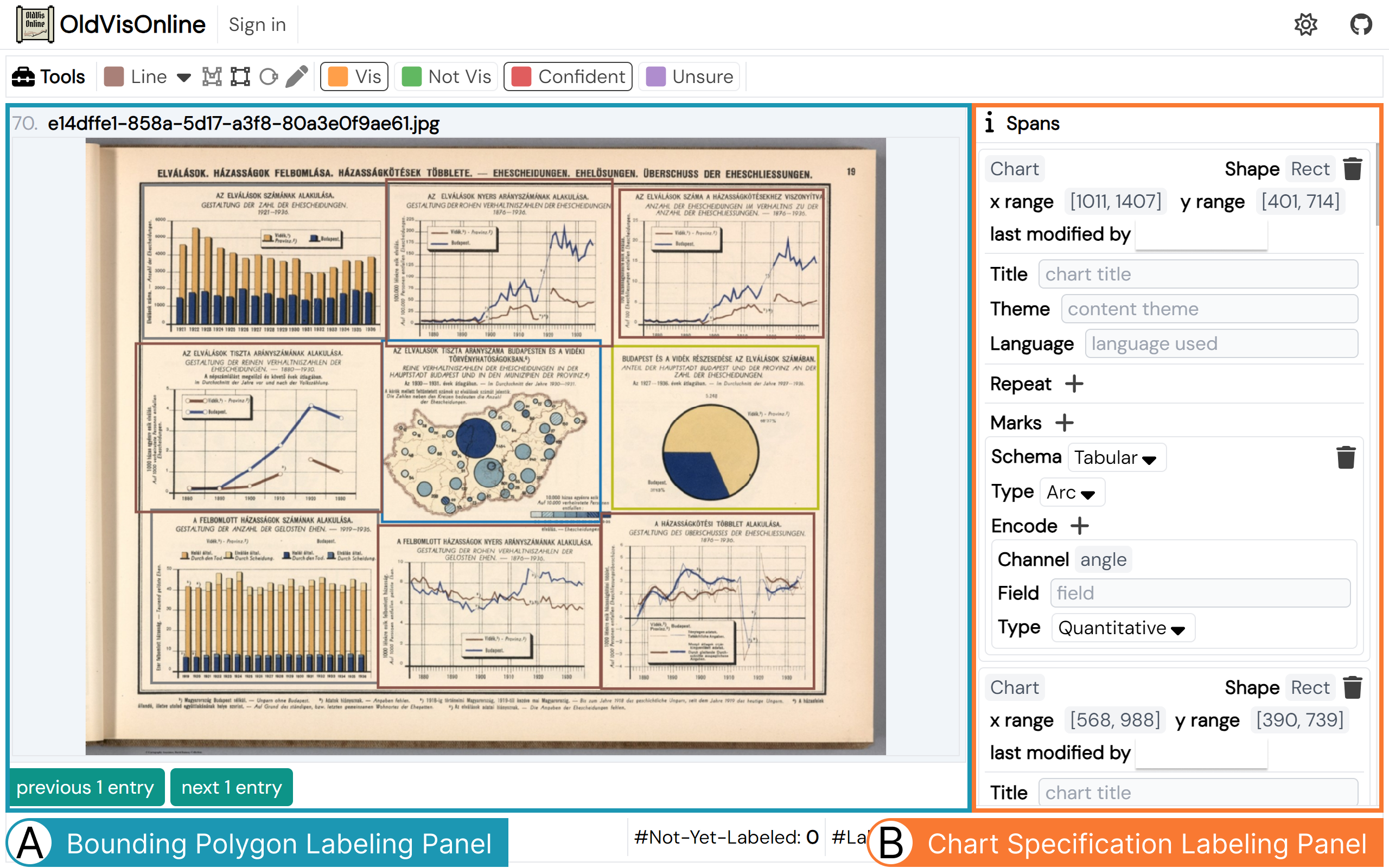}
    \caption{
        \textbf{Annotating visual designs with OldVisOnline's fine-grained labeling interface:}
        (A) The bounding polygon labeling panel where the annotator can annotate the location of charts in a multi-view historical visualization.
        (B) The chart specification labeling panel where the annotator can annotate the visual encoding of the charts.
    }
    \label{fig:fine-grained-labeling-interface}
\end{figure}

To facilitate such usages, OldVisOnline comprises a fine-grained labeling interface (\cref{fig:fine-grained-labeling-interface}).
The interface supports the annotation of bounding polygons of charts within multi-view visualizations (\cref{fig:fine-grained-labeling-interface}(A)).
For each chart, the annotator can annotate the specification of its visual encoding (\cref{fig:fine-grained-labeling-interface}(B)).
We design the specification to resemble the Vega specification~\cite{Satyanarayan2016Reactive}.
The annotator can specify the type of visual elements in the chart and the visual channels used to encode data.
Additional metadata, such as the chart's title, theme, and language, can also be assigned to the charts.
As an initial step towards supporting design space analysis, we have conducted fine-grained annotation for 485 sampled historical visualizations using the labeling interface.

\begin{figure}[!htbp]
    \centering
    \includegraphics[width=\linewidth]{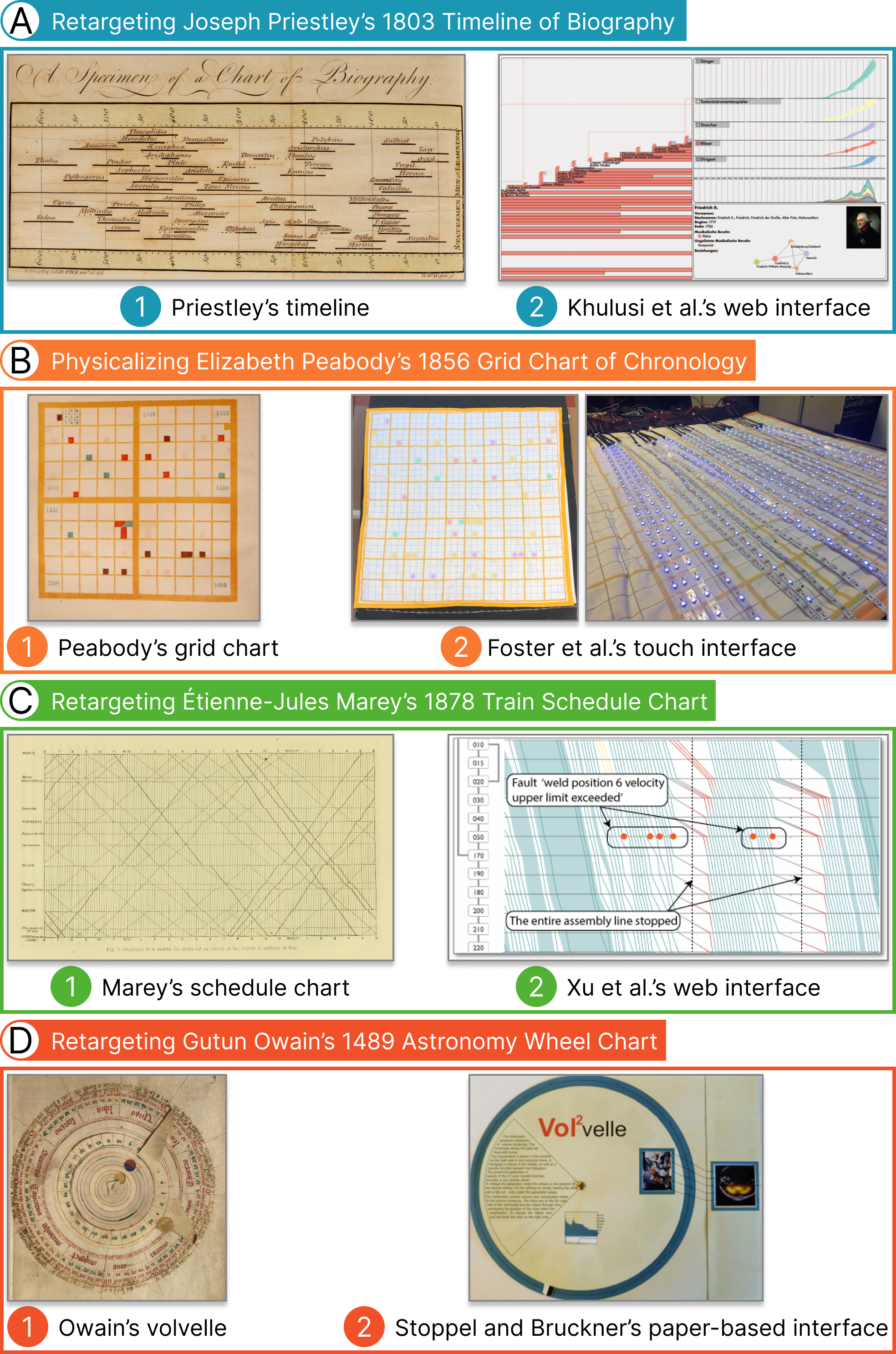}
    \caption{
        \textbf{Reviving historical visualizations:}
        (A1) Joseph Priestley's \href{https://archive.org/details/lecturesonhis1803prie/page/n285}{timeline} published in 1803 (first published in 1764~\cite{Priestley1764Description}) shows historical figures' lifespans~\cite{Priestley1803Lectures}.
        (A2) Khulusi et al. propose an interactive variant of Priestley's timeline as a web interface and retarget the design to visualize musicians' lifespans~\cite{Khulusi2019Interactive}.
        (B1) Elizabeth Peabody's \href{https://archive.org/details/chronologicalhis00peab/page/n25}{grid chart} published in 1856 shows chronological history through glyphs~\cite{Peabody1856Chronological}.
        (B2) Foster et al. recreate Peabody's grid chart through a physicalization implemented with a LED matrix that supports touch interaction~\cite{Foster2017Shape,Klein2022What}.
        (C1) {\'{E}}tienne-Jules Marey's \href{https://archive.org/details/b21919070/page/n49}{schedule chart} published in 1878 shows train arrival and departure times~\cite{Marey1878La}.
        (C2) Xu et al. extend Marey's chart to visualize delays in an assembly line~\cite{Xu2017ViDX}.
        (D1) Gutun Owain's \href{http://hdl.handle.net/10107/4393492\#?cv=12}{volvelle} finished in 1489 shows the position of sun and moon in the zodiac~\cite{Owain1489Hen}.
        (D2) Stoppel and Bruckner use the volvelle design for volume visualization~\cite{Stoppel2017Vol2velle}.
    }
    \label{fig:design}
\end{figure}

\subsection{Inspiration for Contemporary Design}

Historical visualizations are precious samples of the visualization design space.
Many historical visualizations feature unique and insightful designs that may inspire contemporary visualization techniques.

In the following, we give examples of contemporary designs inspired by historical visualizations.
Historical visualizations are usually paper-based presswork that does not support interaction.
An interesting investigation is to enhance historical visualizations with interaction techniques and retarget them to contemporary datasets.

Joseph Priestley's \textit{Chart of Biography} (\cref{fig:design}(A1)) published in 1803 visualizes the lifespans of historical figures~\cite{Priestley1803Lectures}.
Khulusi et al. present an interactive variant of Priestley's \textit{Chart of Biography} and apply it to visualize the lifelines of musicians (\cref{fig:design}(A2))~\cite{Khulusi2019Interactive}.
Brehmer et al. take \textit{Chart of Biography} as a sample to inform the analysis of timeline design space and the development of a timeline authoring tool~\cite{Brehmer2017Timelines}.

Elizabeth Peabody's grid chart (\cref{fig:design}(B1)) published in 1856 visualizes the chronological history where the glyphs in the grids encode historical events~\cite{Peabody1856Chronological}.
Foster et al. recreate Peabody's grid chart digitally and physically~\cite{Foster2017Shape,Klein2022What}.
They demonstrate reviving Peabody's grid chart design as a web-based visualization as well as a physicalization implemented with a LED matrix (\cref{fig:design}(B2)).

{\'{E}}tienne-Jules Marey's schedule chart (\cref{fig:design}(C1)) published in 1878 visualizes the arrival and departure times of trains between Paris and Lyon~\cite{Marey1878La}.
Xu et al. extend Marey's chart to visualize the delay it took products to go through each workstation on an assembly line~\cite{Xu2017ViDX}.
Compared with Marey's original design, Xu et al. aggregates polylines into bands to reduce visual clutter (\cref{fig:design}(C2)).

Gutun Owain's volvelle (\cref{fig:design}(D1)) finished in 1489 can be used to compute the position of the sun and moon in the zodiac~\cite{Owain1489Hen}.
Volvelle, also known as wheel chart, is a paper-based interactive visualization consisting of rotating paper disks.
Owain's chart belongs to the minority of historical visualizations that are interactive.
Following the volvelle design, Stoppel and Bruckner propose an authoring tool to create paper-based interactive volume visualizations (\cref{fig:design}(D2))~\cite{Stoppel2017Vol2velle}.

\subsection{Metadata Analysis}

Valuable findings may be obtained by analyzing the metadata of a large-scale visualization dataset.
For example, we observe some highly productive authors in the \emph{Authors} panel of the image gallery (\cref{fig:online-gallery}(A)), which motivates us to investigate their authorship pattern.
Some authors, such as Henry Gannett (922 entries) and Lajos I. Illyefalvi (439 entries), contribute hundreds of entries to our dataset.
By comparison, some other authors' works are much less indexed in our dataset, including some notable names in the visualization history, such as John Snow (9 entries) and Florence Nightingale (6 entries).

Looking into the indexed works of highly productive authors, we observe that these authors are typically highly involved in producing official statistics.
Many of Henry Gannett's works have shared authorship with the United States Census Office, where he worked.
Similarly, all the works of Lajos I. Illyefalvi have shared authorship with Budapest F{\"{o}}v{\'{a}}ros Statisztikai Hivatal (Capital City Statistical Office, Budapest).

While some highly productive authors have well-known contributions to visualization, such as Henry Gannett's contribution to parallel coordinates, other productive authors are less mentioned.
It is worthwhile to investigate such cases, which may unearth previously unknown treasures and improve our understanding of the development of the visualization subject from a historical perspective.

The above brief example of analyzing the authorship pattern suggests that the metadata analysis of our dataset may bring interesting findings.
Meanwhile, we want to warn users of the potential sample bias in our dataset.
The historical visualizations that survive to date may be a biased sample of the visualizations ever created.
The historical visualizations accessible in digital libraries may be a biased sample of the historical visualizations that survive to date.
Due to these inevitable sources of sample bias, users of our dataset should be cautious when interpreting the metadata statistics.
We recommend that the users use the metadata statistics as inspirations for further analysis rather than as direct conclusions.

\subsection{Metadata Research}

While studies on historical visualizations may rely on the metadata, in some cases, the metadata may not be readily available or needs to be validated.
A related research theme in historical geography is to date the creation time of historical maps from the map content~\cite{Spaeth2000Dating}.
Various map features can be used to infer the time ranges of creation, such as the author information, seals on the map, the toponyms, the naming taboo, the water system, and the admin boundaries.
The creation time of other maps with similar designs may also inform the creation time of the concerned map.

More generally, there is a need to support research and validation of historical visualizations' metadata, including and not limited to creation time.
On the one hand, our dataset will benefit from metadata research, which helps to improve the metadata quality.
On the other hand, our dataset may serve as a corpus for historians to identify historical visualizations similar to their concerned visualization, which may assist metadata research.

   \section{Discussion}
\label{sec:discussion}

In the following, we reflect on the challenges and lessons learned in the dataset construction process.
We also discuss the limitations of our work and the potential future work.

\textbf{Handling API differences:}
While there have been efforts to unify the APIs of digital libraries, such as the International Image Interoperability Framework~\cite{IIIF2011International}, we found in practice that the data sources had different API and metadata designs.
Thus, data wrangling needs to be conducted data source by data source to map the obtained raw metadata into a uniform data structure.
This challenge is also highlighted by Battle et al. when mining SVG-based visualizations from the web~\cite{Battle2018Beagle}.

\textbf{Supporting data structure design iterations:}
Before arriving at the metadata schema introduced in \cref{fig:online-gallery}(C), the data structure design experienced several iterations.
We want to avoid refetching data from the data sources when we add a new metadata field in the iterations.
Thus, we cache the raw metadata obtained from the data sources with the following schema.
All the information obtained from the data sources, whether currently used for generating the metadata or not, is stored in the \inlinecode{sourceData} field of \inlinecode{RawMetadataEntry}.

\begin{lstlisting}[language=TypeScript]
class RawMetadataEntry {
  /** The uuid of the metadata entry. */
  uuid: string
  /** The URL where the raw metadata were fetched. */
  url: string
  /** The name of the data source. */
  source: string
  /** The identifier used by the data source. */
  idInSource: string
  /** The time the entry was saved. */
  accessDate: string
  /** The raw metadata fetched from the URL. */
  sourceData: unknown
}
\end{lstlisting}

Likewise, the data labels introduced in \cref{sec:data-labeling-and-filtering} are stored separately from the metadata.
The data labels and metadata are cross-linked through UUID.
By decoupling the storage of labels and metadata, when updating the labels or the metadata, the other one can be kept untouched, which reduces the complexity of data version control.

\textbf{Improving label quality:}
One of the authors conducted data labeling and quality assurance.
In the future, we plan to improve label quality by majority voting with more expert annotations.
After publicizing the historical visualization dataset and OldVisOnline, we expect to attract more users with visualization expertise to contribute feedback on erroneous labels.

\textbf{Enlarging the dataset:}
An ample scope to further our effort is to enlarge our historical visualization dataset on the following aspects:

\begin{itemize}[leftmargin=3.5mm]
    \item \textbf{Data sources:}
          In addition to the \numDataSources data sources introduced in \cref{sec:data-sources}, we plan to cover more data sources, such as \dbname{Wikimedia Commons}~\cite{WikimediaFoundation2004Wikimedia} and \dbname{Digital Public Library of America}~\cite{DPLA2013Digital}, to improve our coverage of historical visualizations.

    \item \textbf{Search queries:}
          The queries introduced in \cref{sec:search-strategies} and
          \iflabelexists{sec:unify-raw-metadata}{\cref{sec:unify-raw-metadata}}{the supplemental material}
          likely have not harvested all the historical visualizations from the data sources.
          We plan to try more search queries to utilize these data sources fully.

    \item \textbf{Metadata fields:}
          The raw metadata we obtained from the data sources sometimes have richer information than the current metadata schema (\cref{fig:online-gallery}(C)).
          For example, \dbname{David Rumsey Map Collection} and \dbname{Gallica} record the physical size of some collections.
          We plan to investigate the usage scenarios of such unexplored information and work on cleaning and integrating it into our dataset.

\end{itemize}

\textbf{Inferring additional semantic information:}
More usage scenarios of the historical visualization dataset may be facilitated by extracting from each visualization additional information such as:

\begin{itemize}[leftmargin=3.5mm]
    \item \textbf{Bounding boxes and specifications:}
        As discussed in \cref{sec:design-space-analysis}, locations of charts in multi-view visualizations and chart specifications may support fine-grained design space analysis.
        Image segmentation models can be integrated into the labeling interface in \cref{fig:fine-grained-labeling-interface} to reduce the cost of bounding box annotation.
    
    \item \textbf{Embedded text:}
        OCR techniques can be applied to historical visualizations to extract textual information and categorize text by roles (e.g., background information or chart caption).
        The extracted text can be used for content-based searching of the visualization images.

    \item \textbf{Semantic tags:}
        The \inlinecode{tags} of the visualization may capture the visual design (e.g., ``bar chart'' or ``line chart'') and the theme (e.g., ``economics'' or ``demography'').    
        Meanwhile, the data sources do not always provide informative tags.
        In such cases, semantic tags can be extracted from the visualization image and metadata.
        For example, multi-modal models, such as CLIP~\cite{Radford2021Learning}, can be used to compute the embedding of the image and metadata.
        Tags can then be matched by embedding similarity.

    \item \textbf{Relations:}
        The relations between visualizations can be extracted to support in-depth analysis.
        For example, the image content, publisher, and publish date can be used to infer whether two entries are different copies of the same visualizations, which provides candidates for textual criticism studies discussed in \cref{sec:textual-criticism}.
\end{itemize}

\textbf{Extracting entities other than visualizations:}
Visualizations are the concerned entities in our current dataset.
A potential extension is to extract and clean other types of entities, such as \emph{author} and \emph{book}.
Extracting other types of entities may facilitate additional bibliographic analysis tasks, such as coauthorship network analysis.

\begin{figure}[!htbp]
    \centering
    \includegraphics[width=\linewidth]{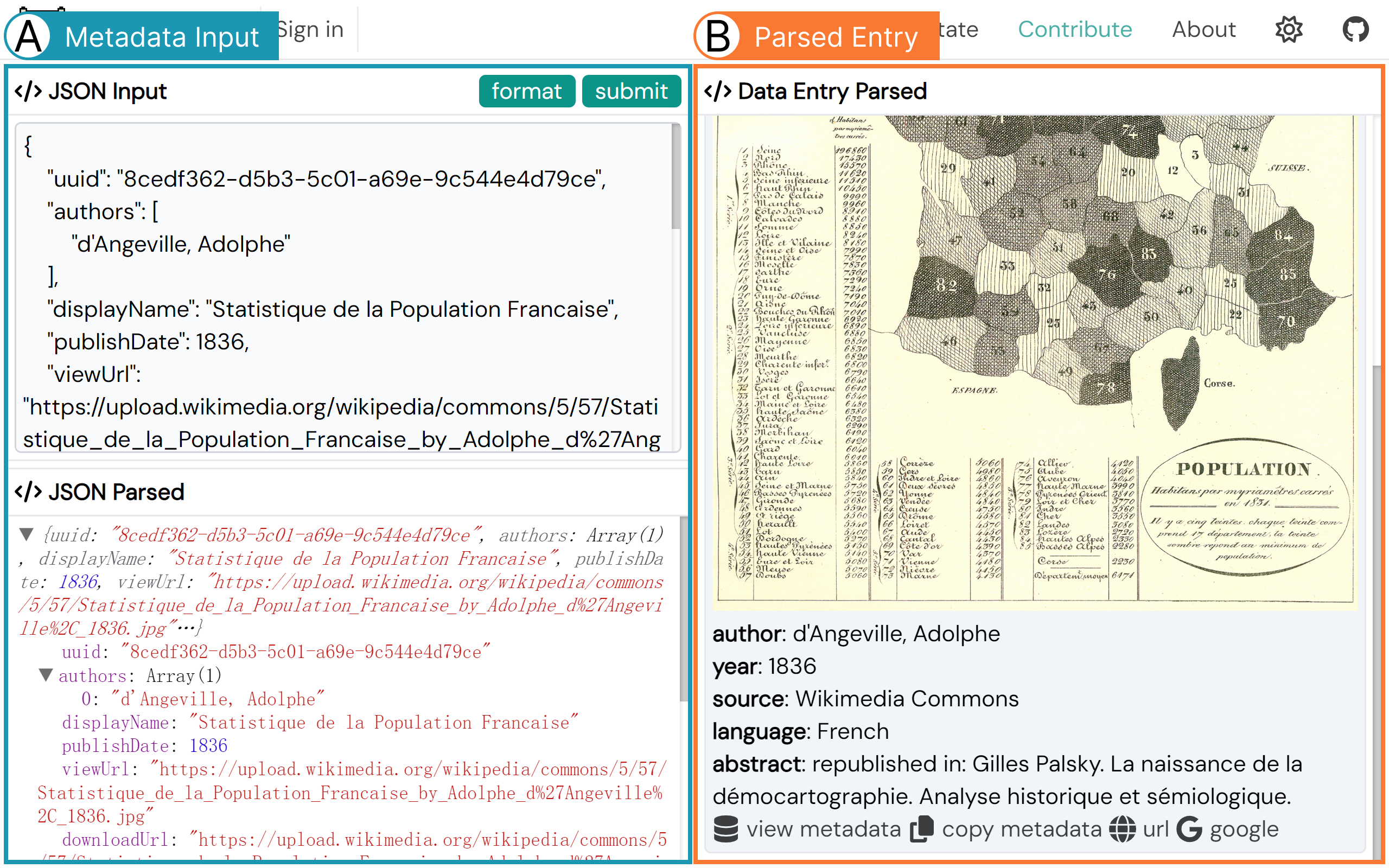}
    \caption{
        \textbf{Contributing historical visualization entries in OldVisOnline:}
        (A) The metadata entry contributed and the parsed JSON object.
        (B) A preview of the parsed metadata entry.
    }
    \label{fig:contribute}
\end{figure}

\textbf{Towards a continuously growing dataset:}
Some of the data sources we use are continuously updated.
For example, when this project was initiated in 2017, \dbname{David Rumsey Map Collection} returned around \num{2800} images for the query \href{https://www.davidrumsey.com/luna/servlet/as/search?q=subject=data+visualization}{subject=data+visualization}.
As of March 29, 2023, the same query returned \num{5412} images, which has almost doubled.
Thus, we see data collection as a continuous process.
The maintainer can carry out the dataset construction workflow introduced in \cref{sec:dataset} periodically to grow the dataset continuously.
Additionally, users may directly contribute historical visualizations by providing the metadata in JSON, as shown in \cref{fig:contribute}.

\textbf{Improving data exploration support:}
This paper focuses on collecting and indexing historical visualizations.
For the next step, data exploration support can be improved to better facilitate the usage scenarios discussed in \cref{sec:usage-scenarios}.
For example, image visualization techniques, such as PhotoMesa~\cite{Bederson2001PhotoMesa}, may be adapted to provide an overview of the image collection.
Additionally, textual criticism described in \cref{sec:textual-criticism} relies on visual comparison.
Image feature extraction and pattern matching may support efficient visual comparison, as demonstrated in the ARIES~\cite{Crissaff2018ARIES} and VeCHArt~\cite{Pfluger2020VeCHArt} systems.

   \section{Conclusion}

This paper contributes a dataset of historical visualizations.
Our dataset consists of \numOldVisShort visualization images published before 1950 with corresponding metadata. 
We have introduced our dataset construction workflow and underlying considerations.
Our OldVisOnline system supports the browsing and labeling of historical visualizations.
We have analyzed in-depth usage scenarios of our dataset, including textual criticism, historical data extraction, design space analysis, inspiration for contemporary design, metadata analysis, and metadata research.

Data visualization has a long history that significantly predates the computer age.
While some historical visualizations are well-known and widely used in courses, the majority of historical visualizations remain underexplored.
Through our work in this paper, we strive to contribute to better utilization of the treasure the pioneers have left us.

   \section*{Supplemental Material}

\iflabelexists{sec:unify-raw-metadata}{\Cref{sec:unify-raw-metadata}}{The supplemental material titled ``Unifying Raw Metadata from Digital Libraries''}
describes our strategy to convert the raw metadata from different data sources into a uniform data structure.
It reflects our implementation as of June 20, 2023. 
The latest implementation and documentation are at \url{\githubLink}.

   \acknowledgments{
    The authors thank Aoyu~Wu, Yixing~Zhang, and Siqi~Shen for their feedback on this work.
    We thank the digital libraries for digitizing historical visualizations and making them accessible.
    The project name, ``OldVisOnline'', is inspired by the Old Maps Online~\cite{KTGOld} project.
    This work is supported by NSFC No. 62272012.
}

   \bibliographystyle{abbrv-doi-hyperref}
   
   \bibliography{assets/bibs/papers,assets/bibs/historical-visualizations,assets/bibs/custom}
\fi

\ifx\hideappendix\undefined

   \pagebreak
   \appendix %
   \begin{table*}[!htbp]
    \caption{
        The raw metadata attributes we use to compute \inlinecode{uuid}, \inlinecode{authors}, \inlinecode{displayName}, \inlinecode{publishDate}, \inlinecode{viewUrl}, \inlinecode{downloadUrl}, \inlinecode{languages}, \inlinecode{tags}, \inlinecode{abstract}, and \inlinecode{rights} in the uniform data structure for each data source.
        ``constant'' refers to the case where the value is a constant independent of the raw metadata.
        ``null'' refers to the case where the value cannot be inferred from the information in the raw metadata.
    }
    \label{table:metadata-processing-1}
    \scriptsize
    \scalebox{0.74}{
        \begin{tabular}{llllllll}
            \toprule
            \textbf{Data Source}             & \inlinecode{uuid}                           & \inlinecode{authors}                             & \inlinecode{displayName}                         & \inlinecode{publishDate}                                        & \inlinecode{viewUrl}                        \\
            \midrule
            David Rumsey Map Collection      & \inlinecode{id}                             & \inlinecode{fieldValues}                         & \inlinecode{displayName}                         & \inlinecode{fieldValues}                                        & \inlinecode{id}                             \\
            Internet Archive                 & \inlinecode{files}                          & \inlinecode{creator}                             & \inlinecode{title}, \inlinecode{files}           & \inlinecode{date}                                               & \inlinecode{identifier}, \inlinecode{files} \\
            Gallica                          & \inlinecode{identifier}, \inlinecode{ordre} & \inlinecode{dc:creator}                          & \inlinecode{legend}, \inlinecode{dc:title}       & \inlinecode{dc:date}                                            & \inlinecode{identifier}, \inlinecode{ordre} \\
            Library of Congress              & \inlinecode{id}                             & \inlinecode{contributors}                        & \inlinecode{title}                               & \inlinecode{date}, \inlinecode{created_published}               & \inlinecode{url}                            \\
            Telefact                         & \inlinecode{viewUrl}                        & constant                                         & \inlinecode{viewUrl}                             & \inlinecode{publishDate}                                        & \inlinecode{viewUrl}                        \\
            British Library Collection Items & \inlinecode{images}                         & \inlinecode{Creator}                             & \inlinecode{Full title}                          & \inlinecode{Published}, \inlinecode{Created}, \inlinecode{Date} & \inlinecode{viewUrl}                        \\
            British Library Images Online    & \inlinecode{viewUrl}                        & \inlinecode{Artist/creator}, \inlinecode{Author} & \inlinecode{Caption}, \inlinecode{Title of Work} & \inlinecode{Place and date of production}                       & \inlinecode{viewUrl}                        \\
            Alabama Maps                     & \inlinecode{viewUrl}                        & \inlinecode{mainAuthor}                          & \inlinecode{titleDescription}                    & \inlinecode{date}                                               & \inlinecode{viewUrl}                        \\
            \bottomrule
        \end{tabular}
    }
\end{table*}

\begin{table*}[!htbp]
    \caption{
        \Cref{table:metadata-processing-1} continued.
    }
    \label{table:metadata-processing-2}
    \scriptsize
    \scalebox{0.74}{
        \begin{tabular}{lllllll}
            \toprule
            \textbf{Data Source}             & \inlinecode{downloadUrl}                                                           & \inlinecode{languages}                         & \inlinecode{tags}                       & \inlinecode{abstract}                               & \inlinecode{rights}                                                                \\
            \midrule
            David Rumsey Map Collection      & \inlinecode{urlSize2}, \inlinecode{urlSize4}                                       & \inlinecode{fieldValues}                       & \inlinecode{fieldValues}                & \inlinecode{fieldValues}                            & constant                                                                           \\
            Internet Archive                 & \inlinecode{identifier}, \inlinecode{server}, \inlinecode{dir}, \inlinecode{files} & \inlinecode{title}                             & \inlinecode{subject}                    & \inlinecode{description}                            & \inlinecode{rights}                                                                \\
            Gallica                          & \inlinecode{identifier}, \inlinecode{ordre}                                        & \inlinecode{dc:language}                       & \inlinecode{dc:type}                    & \inlinecode{dc:format}, \inlinecode{dc:description} & \inlinecode{dc:rights}                                                             \\
            Library of Congress              & \inlinecode{image_url}                                                             & \inlinecode{language}                          & \inlinecode{format}, \inlinecode{genre} & \inlinecode{description}, \inlinecode{notes}        & \inlinecode{rights}, \inlinecode{rights_advisory}, \inlinecode{rights_information} \\
            Telefact                         & \inlinecode{downloadUrl}                                                           & constant                                       & null                                    & null                                                & null                                                                               \\
            British Library Collection Items & \inlinecode{images}                                                                & \inlinecode{Language}, \inlinecode{Full title} & null                                    & \inlinecode{shortDescription}                       & \inlinecode{Usage terms}, \inlinecode{Copyright}                                   \\
            British Library Images Online    & \inlinecode{downloadUrl}                                                           & null                                           & null                                    & \inlinecode{Caption}                                & \inlinecode{Shelfmark}                                                             \\
            Alabama Maps                     & \inlinecode{downloadUrl}                                                           & \inlinecode{titleDescription}                  & null                                    & null                                                & null                                                                               \\
            \bottomrule
        \end{tabular}
    }
\end{table*}

\section{Unifying Raw Metadata from Digital Libraries}
\label{sec:unify-raw-metadata}

This supplemental material describes our strategy to convert the raw metadata from different data sources into the uniform data structure, \inlinecodesmall{MetadataEntry}, shown in
\iflabelexists{fig:online-gallery}{\cref{fig:online-gallery}(C)}{main text Fig. 1(C)}.
As described in
\iflabelexists{sec:discussion}{\cref{sec:discussion}}{main text Sec. 6},
the raw metadata are stored as \inlinecodesmall{RawMetadataEntry}.
The following introduces how each field of \inlinecodesmall{MetadataEntry} is computed from \inlinecodesmall{RawMetadataEntry}.
It reflects our implementation as of June 20, 2023.
Our latest implementation can be found at \url{\githubLink}.
Let ``$\gets$'' denote value assignment.

\textbf{Processing fields related to the data source:}
Three fields related to the data source in \inlinecodesmall{MetadataEntry} are directly ported from \inlinecodesmall{RawMetadataEntry}:

\begin{itemize}[leftmargin=3.5mm]
    \item \inlinecodesmall{MetadataEntry.source.name} $\gets$ \inlinecodesmall{RawMetadataEntry.source}.
    \item \inlinecodesmall{MetadataEntry.source.url} $\gets$ \inlinecodesmall{RawMetadataEntry.url}.
    \item \inlinecodesmall{MetadataEntry.source.accessDate} $\gets$\\ \inlinecodesmall{RawMetadataEntry.accessDate}.
\end{itemize}

\textbf{Processing fields related to the image file properties:}
The image file properties in \inlinecodesmall{MetadataEntry} (\inlinecodesmall{md5}, \inlinecodesmall{phash}, \inlinecodesmall{resolution}, and \inlinecodesmall{fileSize}) are computed from the image file downloaded from \inlinecodesmall{MetadataEntry.downloadUrl}.

\textbf{Processing the remaining fields:}
The processing of the remaining fields of \inlinecodesmall{MetadataEntry} depends on the data source.
\Cref{table:metadata-processing-1} and \Cref{table:metadata-processing-2} summarize the raw metadata fields used to compute each metadata field.
For each data source, below provides details on these aspects:
\begin{itemize}[leftmargin=3.5mm]
    \item \textbf{Data Structure:} The data structure of \inlinecodesmall{SourceData}.
    \item \textbf{Data Processing:} The data processing to compute \inlinecodesmall{MetadataEntry}.
    \item \textbf{Queries:} The queries to gather \inlinecodesmall{SourceData}.
\end{itemize}

\subsection{David Rumsey Map Collection}

\subsubsection{Data Structure}

\inlinecodesmall{SourceData} shows the data structure returned from \dbname{David Rumsey Map Collection}'s API\footnote{\url{https://doc.lunaimaging.com/display/V74D/LUNA+API+Documentation} (Accessed on Mar 1, 2023)}.
The fields that we have not used are not shown.
An entry the API returns corresponds to an image.

\begin{lstlisting}[language=TypeScript]
type Fields = 'Type' | 'Subject' | 'Publication Author'
    | 'Pub Date' | 'Pub Title' | 'Pub Type' | 'Note'
class SourceData {
    /** The title. */
    displayName: string
    /** A paragraph describing the entry. */
    description: string
    /** A unique identifier of the raw entry. */
    id: string
    /** Metadata of various fields. */
    fieldValues: Record<Fields, string[]>[]
    /** A URL of the image with low resolution. */
    urlSize2: string
    /** A URL of the image with high resolution. */
    urlSize4?: string
}
\end{lstlisting}

\subsubsection{Data Processing}

The \inlinecodesmall{MetadataEntry} fields are computed as the following.
The variables on the left-hand side refer to the attributes of \inlinecodesmall{MetadataEntry}.
Unless otherwise noted, the variables on the right-hand side refer to the attributes of \inlinecodesmall{SourceData}.

\begin{itemize}[leftmargin=3.5mm]
    \item \inlinecodesmall{uuid} $\gets$ \inlinecodesmall{RawMetadataEntry.uuid}.
        We use \inlinecodesmall{id} as the seed to generate \inlinecodesmall{RawMetadataEntry.uuid}, a version-5 UUID.

    \item \inlinecodesmall{authors}:
        Parsed from \inlinecodesmall{fieldValues}'s entry with key \inlinecodesmall{`Publication Author'}.
        The authors' names sometimes store the dates of birth and death.
        We use regular expressions to match and trim the dates.
    
    \item \inlinecodesmall{displayName} $\gets$ \inlinecodesmall{displayName}.
    
    \item \inlinecodesmall{publishDate}:
        Parsed from \inlinecodesmall{fieldValues}'s entry with key \inlinecodesmall{`Pub Date'}.
    
    \item \inlinecodesmall{viewUrl} $\gets$ \inlinecodesmall{`https://www.davidrumsey.com/luna/servlet/detail}\\\inlinecodesmall{/{id}'}.
    
    \item \inlinecodesmall{downloadUrl} $\gets$ \inlinecodesmall{urlSize4} || \inlinecodesmall{urlSize2}.
    
    \item \inlinecodesmall{languages}:
        Parsed from the language used in the publication title.
        The publication title is parsed from \inlinecodesmall{fieldValues}'s entry with key \inlinecodesmall{`Pub Title'}.
    
    \item \inlinecodesmall{tags}:
        Parsed from \inlinecodesmall{fieldValues}'s entries with key \inlinecodesmall{`Type'} or \inlinecodesmall{`Subject'} or \inlinecodesmall{`Pub Type'}.

    \item \inlinecodesmall{abstract}:
        Parsed from \inlinecodesmall{fieldValues}'s entry with key \inlinecodesmall{`Note'}.

    \item \inlinecodesmall{rights} $\gets$ \inlinecodesmall{`CC BY-NC-SA 3.0'}\footnote{See ``copyright and permissions'' section in \url{https://www.davidrumsey.com/about} (Accessed on June 28, 2023).}.
\end{itemize}

\subsubsection{Queries}

The queries we have used for this data source are:

\begin{itemize}[leftmargin=3.5mm]
    \item \href{https://www.davidrumsey.com/luna/servlet/as/search?q=subject=data+visualization}{subject=data+visualization}
    \item \href{https://www.davidrumsey.com/luna/servlet/as/search?q=subject=statistical}{subject=statistical}
    \item \href{https://www.davidrumsey.com/luna/servlet/as/search?q=type=chart}{type=chart}
    \item \href{https://www.davidrumsey.com/luna/servlet/as/search?q=type=diagram}{type=diagram}
    \item \href{https://www.davidrumsey.com/luna/servlet/as/search?q=data+visualization}{data+visualization}
    \item \href{https://www.davidrumsey.com/luna/servlet/as/search?q=statistical}{statistical}
    \item \href{https://www.davidrumsey.com/luna/servlet/as/search?q=chart}{chart}
    \item \href{https://www.davidrumsey.com/luna/servlet/as/search?q=diagram}{diagram}
    \item \href{https://www.davidrumsey.com/luna/servlet/as/search?q=data}{data}
\end{itemize}

While \dbname{David Rumsey Map Collection} has a dedicated search index for visualization: \href{https://www.davidrumsey.com/luna/servlet/as/search?q=subject=data+visualization}{subject=data+visualization}, we do not solely rely on this query.
After browsing the returned images, we find this query returns images related to but not necessarily belonging to visualization.
Additionally, this query returns only part of the visualization images in the data source.

\subsection{Internet Archive}

\subsubsection{Data Structure}

\inlinecodesmall{SourceData} shows the data structure returned from \dbname{Internet Archive}'s Python API\footnote{\url{https://archive.org/developers/internetarchive/index.html} (Accessed on Mar 1, 2023)}.
The fields that we have not used are not shown.
An entry the API returns corresponds to an image (e.g., a scanned map) or an image collection (e.g., a scanned book).
When the entry corresponds to an image collection, the \inlinecodesmall{files} contains a zip file of JPEG or JP2 images.
In this case, we construct a metadata entry for each image in the zip file.

\begin{lstlisting}[language=TypeScript]
class SourceData {
    /** Related files stored on the server. */
    files: {
        /** The file name. */
        name: string
    }[]
    /** A URL of the server hosting the files. */
    server: string
    /** The server directory containing the files. */
    dir: string
    /** Metadata of the entry. */
    metadata: {
        /** A unique identifier of the raw entry. */
        identifier: string
        /** The title. */
        title: string
        /** The author(s). */
        creator?: string[] | string
        /** The publish date. */
        date?: string[] | string
        /** A paragraph describing the entry. */
        description?: string
        /** Copyright information. */
        rights?: string
        /** Themes that serve as search indices. */
        subject?: string[] | string
    }
}
\end{lstlisting}

\subsubsection{Data Processing}

The \inlinecodesmall{MetadataEntry} fields are computed as the following.
\dbname{Internet Archive}'s Python API supports download for \inlinecodesmall{files} in the query results.
We use \inlinecodesmall{name} to denote the downloaded image filename.
The image is either directly downloaded with the API or unzipped from the zip file downloaded with the API.

\begin{itemize}[leftmargin=3.5mm]
    \item \inlinecodesmall{uuid}:
        We use \inlinecodesmall{name} as the seed to generate a version-5 UUID.
        We do not directly assign \inlinecodesmall{RawMetadataEntry.uuid} to \inlinecodesmall{uuid}, as otherwise, all the images in a zip file would have the same \inlinecodesmall{uuid}.
    
    \item \inlinecodesmall{authors}:
        Parsed from \inlinecodesmall{metadata.creator}.
    
    \item \inlinecodesmall{displayName}:
        \begin{itemize}[leftmargin=3.5mm]
            \item If the entry is constructed from an individual image, \inlinecodesmall{displayName} $\gets$ \inlinecodesmall{metadata.title}.
            
            \item If the entry is constructed from an image in a zip file, we append the page index (denoted as \inlinecodesmall{pageIndex}) to \inlinecodesmall{displayName}.
            \inlinecodesmall{pageIndex} can be parsed from \inlinecodesmall{name}.
        \end{itemize}
    
    \item \inlinecodesmall{publishDate}:
        Parsed from \inlinecodesmall{metadata.date} with regular expressions.
    
    \item \inlinecodesmall{viewUrl}:
        \begin{itemize}[leftmargin=3.5mm]
            \item If the entry is constructed from an individual image,
            \inlinecodesmall{viewUrl} $\gets$ \inlinecodesmall{`https://archive.org/details/{metadata.identifier}'}.
            
            \item If the entry is constructed from an image in a zip file,
            \inlinecodesmall{viewUrl} $\gets$ \inlinecodesmall{`https://archive.org/details/{metadata.identifier}/page}\\\inlinecodesmall{/n{pageIndex}'}.
        \end{itemize}
    
    \item \inlinecodesmall{downloadUrl}:
        \begin{itemize}[leftmargin=3.5mm]
            \item If the entry is constructed from an individual image,
            \inlinecodesmall{downloadUrl} $\gets$ \inlinecodesmall{`https://archive.org/download/{metadata.identifier}/}\\\inlinecodesmall{name'}.
            
            \item If the entry is constructed from an image in a zip file, we compose \inlinecodesmall{downloadUrl} with \inlinecodesmall{server}, \inlinecodesmall{dir}, and \inlinecodesmall{name}.
        \end{itemize}
    
    \item \inlinecodesmall{languages}:
        Parsed from the language used in \inlinecodesmall{metadata.title}.
    
    \item \inlinecodesmall{tags}:
        Parsed from \inlinecodesmall{metadata.subject}.
        The author names are sometimes stored in \inlinecodesmall{metadata.subject}.
        We use a part-of-speech tagging model to detect and discard the author names.

    \item \inlinecodesmall{abstract}:
        Parsed from \inlinecodesmall{metadata.description}, which can be a plain string or an HTML string.
        When it is an HTML string, we convert it into a plain string by removing the HTML tags.

    \item \inlinecodesmall{rights}:
        Parsed from \inlinecodesmall{metadata.rights}.
\end{itemize}

\subsubsection{Queries}

The queries we have used for this data source are:

\begin{itemize}[leftmargin=3.5mm]
    \item \href{https://archive.org/search?query=(statistical+atlas)+AND+-collection:(david-rumsey-map-collection)+AND+date:[0001-01-01+TO+1990-12-31]}{(statistical atlas) AND -collection:(david-rumsey-map-collection) AND date:[0001-01-01 TO 1990-12-31]}
    \item \href{https://archive.org/search?query=creator:(Snow,+John)+AND+-collection:(david-rumsey-map-collection)+AND+date:[1800-01-01+TO+1950-12-31]}{creator:(Snow, John) AND -collection:(david-rumsey-map-collection) AND date:[1800-01-01 TO 1950-12-31]}
    \item \href{https://archive.org/search?query=creator:(Cheysson,+Émile)+AND+-collection:(david-rumsey-map-collection)+AND+date:[0001-01-01+TO+1990-12-31]}{creator:(Cheysson, Émile) AND -collection:(david-rumsey-map-collection) AND date:[0001-01-01 TO 1990-12-31]}
\end{itemize}

\dbname{Internet Archive} has integrated various data sources, including \dbname{David Rumsey Map Collection}\footnote{\url{https://archive.org/details/david-rumsey-map-collection} (Accessed on Mar 1, 202)}.
Meanwhile, compared with data retrieved from \dbname{David Rumsey Map Collection}'s API, \dbname{Internet Archive}'s index of \dbname{David Rumsey Map Collection} images is not up-to-date and stores fewer metadata fields.
Thus, we use \dbname{David Rumsey Map Collection}'s API to query images in it instead of simply using its images readily indexed in \dbname{Internet Archive}.

When querying \dbname{Internet Archive}, we exclude the images from \dbname{David Rumsey Map Collection} to avoid storing the same image twice.
The exclusion corresponds to the query condition ``-collection:(david-rumsey-map-collection)'' in the queries above.

In addition to the queries above, we also trialed other queries but failed to find visualization images when we browsed the query results.
Examples of failed queries include:

\begin{itemize}[leftmargin=3.5mm]
    \item \href{https://archive.org/search?query=(visualization)+AND+mediatype:(image)+AND+-collection:(david-rumsey-map-collection)+AND+date:[0001-01-01+TO+1990-12-31]}{(visualization) AND mediatype:(image) AND -collection:(david-rumsey-map-collection) AND date:[0001-01-01 TO 1990-12-31]}
    \item \href{https://archive.org/search?query=creator:(Minard,+Charles+Joseph)+AND+-collection:(david-rumsey-map-collection)+AND+date:[0001-01-01+TO+1990-12-31]}{creator:(Minard, Charles Joseph) AND -collection:(david-rumsey-map-collection) AND date:[0001-01-01 TO 1990-12-31]}
\end{itemize}

\subsection{Gallica}

\subsubsection{Data Structure}

Unlike the other data sources, \dbname{Gallica}'s API returns an XML string instead of a JSON string.
Thus, we first convert the XML string into a JSON object.
The XML node names convert to keys in the JSON object.
The XML nodes' text values convert to values in the JSON object.
The following focuses on the JSON object after conversion.

\inlinecodesmall{SourceData} shows the data structure returned from \dbname{Gallica}'s search API\footnote{\url{https://api.bnf.fr/api-gallica-de-recherche} (Accessed on Mar 1, 2023)} and metadata API\footnote{\url{https://api.bnf.fr/api-document-de-gallica} (Accessed on Mar 1, 2023)}.
The fields that we have not used are not shown.
Each entry the API returns corresponds to a collection of images stored in \inlinecodesmall{pages}.
We construct a metadata entry for each image in the collection.

\begin{lstlisting}[language=TypeScript]
class SourceData {
    /** A unique identifier of the raw entry. */
    identifier: string
    record: {
        /** The title. */
        'dc:title': string | string[]
        /** The author(s). */
        'dc:creator'?: string | string[]
        /** The publish date. */
        'dc:date'?: string | string[]
        /** Physical and electronic storage format(s). */
        'dc:format'?: string | string[]
        /** The language(s) used in the entry. */
        'dc:language'?: string | string[]
        /** Type(s) of the entry. */
        'dc:type'?: string | string[]
        /** Copyright information. */
        'dc:rights'?: string[]
        /** A paragraph describing the entry. */
        'dc:description'?: string | string[]
    }
    pages: {
        /** The page index. */
        ordre: string
        /** The page title. */
        legend?: string
    }[]
}
\end{lstlisting}

\subsubsection{Data Processing}

The \inlinecodesmall{MetadataEntry} fields are computed as the following.

\begin{itemize}[leftmargin=3.5mm]
    \item \inlinecodesmall{uuid}:
        We use \inlinecodesmall{identifier} and \inlinecodesmall{ordre} as the seed to generate a version-5 UUID.
    
    \item \inlinecodesmall{authors}:
        Parsed from \inlinecodesmall{record[`dc:creator']}.
        The authors' names sometimes store the dates of birth and death.
        We use regular expressions to match and trim the dates.
    
    \item \inlinecodesmall{displayName}:
        If \inlinecodesmall{legend} exists,
        \inlinecodesmall{displayName} $\gets$ \inlinecodesmall{legend}.
        Otherwise \inlinecodesmall{displayName} is parsed from \inlinecodesmall{record[`dc:title']}.
    
    \item \inlinecodesmall{publishDate}:
        Parsed from \inlinecodesmall{record[`dc:date']} with regular expressions.
    
    \item \inlinecodesmall{viewUrl} $\gets$
        \inlinecodesmall{`{identifier}/f{ordre}.item'}.
    
    \item \inlinecodesmall{downloadUrl} $\gets$
        \inlinecodesmall{`{identifier}/f{ordre}.highres'}.
    
    \item \inlinecodesmall{languages}:
        Parsed from \inlinecodesmall{record[`dc:language']}.
    
    \item \inlinecodesmall{tags}:
        Parsed from \inlinecodesmall{record[`dc:type']}.

    \item \inlinecodesmall{abstract}:
        Merges \inlinecodesmall{record[`dc:format']} and \inlinecodesmall{record[`dc:}\\\inlinecodesmall{description']}.

    \item \inlinecodesmall{rights}:
        Parsed from \inlinecodesmall{record[`dc:rights']}.
\end{itemize}

\subsubsection{Queries}

\dbname{Gallica} is the digital library of the national library of France.
We use French search terms as most entries in \dbname{Gallica} are indexed in French.
The queries we have used for this data source are:

\begin{itemize}[leftmargin=3.5mm]
    \item \href{https://gallica.bnf.fr/SRU?operation=searchRetrieve&query=dc.title+all+"carte+figurative"}{dc.title all ``carte figurative''}
        (``carte figurative'' translates to ``figurative map'' or ``figurative chart''. This query returns the same as \href{https://gallica.bnf.fr/SRU?operation=searchRetrieve&query=dc.title+all+"cartes+figurative"}{dc.title all ``cartes figurative''}.)
    
    \item \href{https://gallica.bnf.fr/SRU?operation=searchRetrieve&query=dc.title+all+"tableau+graphique"}{dc.title all ``tableau graphique''}
    
    \item \href{https://gallica.bnf.fr/SRU?operation=searchRetrieve&query=dc.title+all+"statistique+graphique"}{dc.title all ``statistique graphique''}
    
    \item \href{https://gallica.bnf.fr/SRU?operation=searchRetrieve&query=dc.title+all+"atlas+de"}{dc.title all ``atlas de''}
        (This query returns the same as \href{https://gallica.bnf.fr/SRU?operation=searchRetrieve&query=dc.title+all+"atlas"}{dc.title all ``atlas''}.)
\end{itemize}

We decided not to use ``carte'' (can be translated to ``chart'') and ``graphique'' (can be translated to ``graph'') as search terms because they are polysemes.
The majority of their query results are irrelevant.

\subsection{Library of Congress}

\subsubsection{Data Structure}

\inlinecodesmall{SourceData} shows the data structure returned from \dbname{Library of Congress}'s search API\footnote{\url{https://www.loc.gov/apis/} (Accessed on Mar 1, 2023)}.
The fields that we have not used are not shown.
An entry the API returns corresponds to an image.

\begin{lstlisting}[language=TypeScript]
class SourceData {
    /** The HTTP (instead of HTTPS) URL of the entry. */
    id: string
    /** URLs of the image with different resolutions. */
    image_url: string[]
    /** The information for displayed on loc.gov. */
    item: {
        /** The author(s). */
        contributors?: string[]
        /** The publish date. */
        created_published?: string | string[]
        /** The entry format. */
        format?: string | string[]
        /** The entry genre. */
        genre?: string[]
        /** The notes on the entry. */
        notes?: string[]
        /** Copyright information. */
        rights?: string
        rights_advisory?: string | string[]
        rights_information?: string
    }
    /** The title. */
    title: string
    /** The entry URL on the loc.gov website. */
    url: string
    /** The publish date. */
    date?: string
    /** Paragraphs describing the entry. */
    description?: string[]
    /** The languages used in the entry. */
    language?: string[]
}
\end{lstlisting}

\subsubsection{Data Processing}

The \inlinecodesmall{MetadataEntry} fields are computed as the following.

\begin{itemize}[leftmargin=3.5mm]
    \item \inlinecodesmall{uuid} $\gets$ \inlinecodesmall{RawMetadataEntry.uuid}.
        We use \inlinecodesmall{id} as the seed to generate \inlinecodesmall{RawMetadataEntry.uuid}, a version-5 UUID.

    \item \inlinecodesmall{authors} $\gets$ \inlinecodesmall{item.contributors}.
    
    \item \inlinecodesmall{displayName} $\gets$ \inlinecodesmall{title}.
    
    \item \inlinecodesmall{publishDate}:
        Parsed from \inlinecodesmall{date} and \inlinecodesmall{item.created_published} with regular expressions.
    
    \item \inlinecodesmall{viewUrl} $\gets$ \inlinecodesmall{url}.
    
    \item \inlinecodesmall{downloadUrl} $\gets$ \inlinecodesmall{image_url[-1]}.
        The last entry of \inlinecodesmall{image_url} is the version of the image with the highest resolution.
    
    \item \inlinecodesmall{languages}:
        Parsed from \inlinecodesmall{language}.
    
    \item \inlinecodesmall{tags}:
        Parsed by merging \inlinecodesmall{item.format} and \inlinecodesmall{item.genre}.

    \item \inlinecodesmall{abstract}:
        If \inlinecodesmall{description} exists, \inlinecodesmall{abstract} is computed by concatenating all its entries.
        Otherwise, \inlinecodesmall{abstract} is computed by concatenating all the entries of \inlinecodesmall{item.notes}.

    \item \inlinecodesmall{rights}:
        Parsed from \inlinecodesmall{item.rights}, \inlinecodesmall{item.rights_advisory}, and \inlinecodesmall{item.rights_information}.
\end{itemize}

\subsubsection{Queries}

The queries we have used for this data source are:

\begin{itemize}[leftmargin=3.5mm]
    \item \href{https://www.loc.gov/maps/?fo=json&fa=online-format:image}{maps/?}
    \item \href{https://www.loc.gov/photos/?fo=json&fa=online-format:image&q=chart}{photos/?q=chart}
    \item \href{https://www.loc.gov/photos/?fo=json&fa=online-format:image&q=diagram}{photos/?q=diagram}
    \item \href{https://www.loc.gov/photos/?fo=json&fa=online-format:image&q=visualization}{photos/?q=visualization}
    \item \href{https://www.loc.gov/photos/?fo=json&fa=online-format:image&q=graph}{photos/?q=graph}
    \item \href{https://www.loc.gov/photos/?fo=json&fa=online-format:image&q=statistic}{photos/?q=statistic}
    \item \href{https://www.loc.gov/photos/?fo=json&fa=online-format:image&q=popular+and+applied+graphic+art}{photos/?q=popular+and+applied+graphic+art}
\end{itemize}

\subsection{Telefact}

\subsubsection{Data Structure}

\dbname{Telefact: 1938-1945 by Pictograph Corporation} is a digital collection\footnote{\url{https://modley-telefact-1939-1945.tumblr.com/} (Accessed on Mar 1, 2023)} curated by Jason Forrest on Rudolf Modley's ISOTYPE and pictographs.
The digital curation is hosted as a blog without an official API.
We use HTTP GET requests to fetch the webpages as HTML strings and parse the relevant information with regular expressions.
\inlinecodesmall{SourceData} shows the data structure of our parsing result.
A parsed entry corresponds to an image.

\begin{lstlisting}[language=TypeScript]
class SourceData {
    /** The publish date parsed from the webpage. */
    publishDate: {
        year: number
        month: number
        day: number
    }
    /** The webpage URL. */
    viewUrl: string
    /** The image URL parsed from the webpage. */
    downloadUrl: string
}
\end{lstlisting}

\subsubsection{Data Processing}

The \inlinecodesmall{MetadataEntry} fields are computed as the following.

\begin{itemize}[leftmargin=3.5mm]
    \item \inlinecodesmall{uuid} $\gets$ \inlinecodesmall{RawMetadataEntry.uuid}.
        We use \inlinecodesmall{viewUrl} as the seed to generate \inlinecodesmall{RawMetadataEntry.uuid}, a version-5 UUID.

    \item \inlinecodesmall{authors} $\gets$ \inlinecodesmall{[`Modley, Rudolf']}.
    
    \item \inlinecodesmall{displayName}:
        Parsed from \inlinecodesmall{viewUrl}.
    
    \item \inlinecodesmall{publishDate} $\gets$ \inlinecodesmall{publishDate}.
    
    \item \inlinecodesmall{viewUrl} $\gets$ \inlinecodesmall{viewUrl}.
    
    \item \inlinecodesmall{downloadUrl} $\gets$ \inlinecodesmall{downloadUrl}.
    
    \item \inlinecodesmall{languages} $\gets$ \inlinecodesmall{[`eng']}.
\end{itemize}

\inlinecodesmall{tags}, \inlinecodesmall{abstract}, and \inlinecodesmall{rights} cannot be inferred from the information in the data source and are thus left empty.

\subsubsection{Queries}

All the images in this data source are collected.

\subsection{British Library}

\dbname{British Library} provides multiple online galleries, including \dbname{Collection Items} and \dbname{Images Online}.
The online galleries do not have official APIs.
We use HTTP GET requests to fetch the webpages as HTML strings and parse the relevant information from the DOM.
The following two sections introduce our strategy to construct \inlinecodesmall{MetadataEntry} for \dbname{Collection Items} and \dbname{Images Online}.

\subsection{British Library Collection Items}

\subsubsection{Data Structure}

\inlinecodesmall{SourceData} shows the data structure we parsed from the \dbname{Collection Items} website\footnote{\url{https://www.bl.uk/collection-items} (Accessed on Mar 1, 2023)}.
A parsed entry corresponds to a collection of images stored as URLs in \inlinecodesmall{images}.
We construct a metadata entry for each image in the collection.

The images in different image collections sometimes overlap.
We merge the metadata when an image is stored more than once.
In the following, we use \inlinecodesmall{image} (i.e., the image URL) to denote an entry in \inlinecodesmall{images}.

\begin{lstlisting}[language=TypeScript]
class SourceData {
    item: {
        /** The creation date. */
        Created?: string
        /** The author(s). */
        Creator?: string
        /** The usage terms. */
        'Usage terms'?: string
        /** A related date. */
        Date?: string
        /** The title. */
        'Full title'?: string
        /** The language used in the entry. */
        Language?: string
        /** The publish date. */
        Published?: string
        /** A paragraph describing the entry. */
        shortDescription?: string
        /** Copyright information. */
        Copyright?: string
    }
    /** The webpage URL. */
    viewUrl: string
    /** The image URLs. */
    images?: string[]
}
\end{lstlisting}

\subsubsection{Data Processing}

The \inlinecodesmall{MetadataEntry} fields are computed as the following.

\begin{itemize}[leftmargin=3.5mm]
    \item \inlinecodesmall{uuid}:
        We use \inlinecodesmall{image} as the seed to generate a version-5 UUID.
    
    \item \inlinecodesmall{authors}:
        Parsed by segmenting \inlinecodesmall{Creator}.
    
    \item \inlinecodesmall{displayName} $\gets$ \inlinecodesmall{item[`Full title']}.
    
    \item \inlinecodesmall{publishDate}:
        Parsed from \inlinecodesmall{item.Published}, \inlinecodesmall{item.Created}, and \inlinecodesmall{item.Date} with regular expressions.
    
    \item \inlinecodesmall{viewUrl} $\gets$ \inlinecodesmall{viewUrl}.
    
    \item \inlinecodesmall{downloadUrl} $\gets$ \inlinecodesmall{image}.
    
    \item \inlinecodesmall{languages}:
        If \inlinecodesmall{item.Language} exists, \inlinecodesmall{languages} is computed by segmenting \inlinecodesmall{item.Language} and converting segments to standard ISO 639-3 language codes.
        Otherwise, \inlinecodesmall{languages} is parsed from the language used in \inlinecodesmall{item[`Full title']}.
    
    \item \inlinecodesmall{abstract} $\gets$ \inlinecodesmall{item.shortDescription}.
    
    \item \inlinecodesmall{rights} $\gets$ \inlinecodesmall{item[`Usage terms']} || \inlinecodesmall{item[`Copyright']}.
\end{itemize}

\inlinecodesmall{tags} cannot be inferred from the information in the data source and is thus left empty.

\subsubsection{Queries}

All the images in this data source are collected.

\subsection{British Library Images Online}

\subsubsection{Data Structure}

\inlinecodesmall{SourceData} shows the data structure we parsed from the \dbname{Images Online} website\footnote{\url{https://imagesonline.bl.uk} (Accessed on Mar 1, 2023)}.
A parsed entry corresponds to an image.

\begin{lstlisting}[language=TypeScript]
class SourceData {
    /** The webpage URL. */
    viewUrl: string
    /** The image URL parsed from the webpage. */
    downloadUrl: string
    /** The caption. */
    Caption?: string
    /** The title. */
    'Title of Work'?: string
    /** The shelfmark of the physical storage. */
    Shelfmark: string
    /** The publish address and date. */
    'Place and date of production'?: string
    /** The creator(s). */
    'Artist/creator'?: string
    /** The author(s). */
    Author?: string
}
\end{lstlisting}

\subsubsection{Data Processing}

The \inlinecodesmall{MetadataEntry} fields are computed as the following.

\begin{itemize}[leftmargin=3.5mm]
    \item \inlinecodesmall{uuid} $\gets$ \inlinecodesmall{RawMetadataEntry.uuid}.
        We use \inlinecodesmall{viewUrl} as the seed to generate \inlinecodesmall{RawMetadataEntry.uuid}, a version-5 UUID.

    \item \inlinecodesmall{authors}:
        Parsed by segmenting \inlinecodesmall{`Artist/creator'} and \inlinecodesmall{Author}.
    
    \item \inlinecodesmall{displayName} $\gets$ \inlinecodesmall{`Title of Work'} || \inlinecodesmall{Caption}.
    
    \item \inlinecodesmall{publishDate}:
        Parsed from \inlinecodesmall{`Place and date of production'} with regular expressions.
    
    \item \inlinecodesmall{viewUrl} $\gets$ \inlinecodesmall{viewUrl}.
    
    \item \inlinecodesmall{downloadUrl} $\gets$ \inlinecodesmall{downloadUrl}.
    
    \item \inlinecodesmall{abstract} $\gets$ \inlinecodesmall{Caption}.
    
    \item \inlinecodesmall{rights} $\gets$ \inlinecodesmall{`Copyright British Library Board (Shelfmark:}\\\inlinecodesmall{{Shelfmark})'}\footnote{
        See the ``Free download'' section in \url{https://imagesonline.bl.uk/} (Accessed on June 28, 2023).
    }.
\end{itemize}

\inlinecodesmall{languages} and \inlinecodesmall{tags} cannot be inferred from the information in the data source and are thus left empty.

\subsubsection{Queries}

\dbname{British Library Images Online} has no official API.
Its website supports searching.
The search terms we have used are:

\begin{itemize}[leftmargin=3.5mm]
    \item \href{https://imagesonline.bl.uk/search/?searchQuery=chart}{chart}
    \item \href{https://imagesonline.bl.uk/search/?searchQuery=map}{map}
\end{itemize}

\subsection{Alabama Maps}

\subsubsection{Data Structure}

\dbname{Alabama Maps} is a digital collection of maps from various data sources, including the University of Alabama Map Library.
We focus on the Historical Map Archive\footnote{\url{http://alabamamaps.ua.edu/historicalmaps/} (Accessed on Mar 1, 2023)} of \dbname{Alabama Maps}.
The digital collection has no official API.
We use HTTP GET requests to fetch the webpages as HTML
strings and parse the relevant information from the DOM.
\inlinecodesmall{SourceData} shows the data structure of our parsing result.
A parsed entry corresponds to an image.

\begin{lstlisting}[language=TypeScript]
class SourceData {
    /** The author(s). */
    mainAuthor?: string
    /** The title. */
    titleDescription: string
    /** The publish date. */
    date: string
    /** The webpage URL. */
    viewUrl: string
    /** The image URL. */
    downloadUrl: string
}
\end{lstlisting}

\subsubsection{Data Processing}

The \inlinecodesmall{MetadataEntry} fields are computed as the following.

\begin{itemize}[leftmargin=3.5mm]
    \item \inlinecodesmall{uuid} $\gets$ \inlinecodesmall{RawMetadataEntry.uuid}.
        We use \inlinecodesmall{viewUrl} as the seed to generate \inlinecodesmall{RawMetadataEntry.uuid}, a version-5 UUID.

    \item \inlinecodesmall{authors}:
        Parsed by segmenting \inlinecodesmall{mainAuthor}.
    
    \item \inlinecodesmall{displayName} $\gets$ \inlinecodesmall{titleDescription}.
    
    \item \inlinecodesmall{publishDate}:
        Parsed from \inlinecodesmall{date} with regular expressions.
    
    \item \inlinecodesmall{viewUrl} $\gets$ \inlinecodesmall{viewUrl}.
    
    \item \inlinecodesmall{downloadUrl} $\gets$ \inlinecodesmall{downloadUrl}.
    
    \item \inlinecodesmall{languages}:
        Parsed from the language used in \inlinecodesmall{titleDescription}.
\end{itemize}

\inlinecodesmall{tags}, \inlinecodesmall{abstract}, and \inlinecodesmall{rights} cannot be inferred from the information in the data source and are thus left empty.

\subsubsection{Queries}

\dbname{Alabama Maps} has no official API.
Its website does not provide a search function.
Instead, it categorizes images by themes such as ``National Forests'' and ``World War I''.
The categories we have processed are:

\begin{itemize}[leftmargin=3.5mm]
    \item \href{http://alabamamaps.ua.edu/historicalmaps/worldwarI}{worldwarI}
    \item \href{http://alabamamaps.ua.edu/historicalmaps/civilwar/}{civilwar}
\end{itemize}

The Historical Map Archive of \dbname{Alabama Maps} contains 6 image categories.
After browsing the images in the 5 categories titled
``ALABAMA'', ``THE UNITED STATES AND CANADA'', ``THE WORLD'', ``WESTERN HEMISPHERE'', and ``EASTERN HEMISPHERE'',
we failed to find visualizations.
``SPECIAL TOPICS'' is the only category that we found visualizations.

Among the 17 subcategories in ``SPECIAL TOPICS'', the webpage of 7 subcategories
(``Native Americans'',  ``American Revolution'', ``Biblical'', ``Mississippi River''
``Mexican American War'', ``World War I'', and ``American Civil War'')
have similar HTML DOM structures.
The webpages of the other 10 subcategories have different DOM structures, which makes it hard to parse their metadata.
Thus, we decided to focus on these 7 subcategories for now.

Upon reviewing the images within the 7 subcategories, we found that the images in 5 subcategories are mainly conventional geographic maps outside our scope.
Therefore, we only retrieved images from two subcategories: ``World War I'' and ``American Civil War''.

\fi

\end{document}